\documentclass{pasj00}
\Received{2013 May 9}
\Accepted{2014 October 6}
\SetRunningHead{M. Konishi et al.}{Atacama Near-Infrared Camera for miniTAO telescope}

\usepackage{graphicx}
\usepackage{color}

\usepackage{times}

\begin{document}

\title{ANIR : Atacama Near-Infrared Camera for the 1.0-m miniTAO Telescope}
\author{
        Masahiro~\textsc{Konishi},\altaffilmark{1}
        Kentaro~\textsc{Motohara},\altaffilmark{1}
        Ken~\textsc{Tateuchi},\altaffilmark{1}
        Hidenori~\textsc{Takahashi},\altaffilmark{1}
        Yutaro~\textsc{Kitagawa},\altaffilmark{1}
        Natsuko~\textsc{Kato},\altaffilmark{1}
        Shigeyuki~\textsc{Sako},\altaffilmark{1}
        Yuka K.~\textsc{Uchimoto},\altaffilmark{1,2}
        Koji~\textsc{Toshikawa},\altaffilmark{1}
        Ryou~\textsc{Ohsawa},\altaffilmark{3}
        Tomoyasu~\textsc{Yamamuro},\altaffilmark{4}
        Kentaro~\textsc{Asano},\altaffilmark{1}
        Yoshifusa~\textsc{Ita},\altaffilmark{6,2}
        Takafumi~\textsc{Kamizuka},\altaffilmark{1}
        Shinya~\textsc{Komugi},\altaffilmark{8,9,6}
        Shintaro~\textsc{Koshida},\altaffilmark{1,5}
        Sho~\textsc{Manabe},\altaffilmark{11}
        Noriyuki~\textsc{Matsunaga},\altaffilmark{10,3}
        Takeo~\textsc{Minezaki},\altaffilmark{1}
        Tomoki~\textsc{Morokuma},\altaffilmark{1}
        Asami~\textsc{Nakashima},\altaffilmark{3,6,7}
        Toshinobu~\textsc{Takagi},\altaffilmark{8}
        Toshihiko~\textsc{Tanab{\'e}},\altaffilmark{1}
        Mizuho~\textsc{Uchiyama},\altaffilmark{1}
        Tsutomu~\textsc{Aoki},\altaffilmark{10}
        Mamoru~\textsc{Doi},\altaffilmark{1}
        Toshihiro~\textsc{Handa},\altaffilmark{1,12}
        Daisuke~\textsc{Kato},\altaffilmark{3,13}
        Kimiaki~\textsc{Kawara},\altaffilmark{1}
        Kotaro~\textsc{Kohno},\altaffilmark{1}
        Takashi~\textsc{Miyata},\altaffilmark{1}
        Tomohiko~\textsc{Nakamura},\altaffilmark{1,3}
        Kazushi~\textsc{Okada},\altaffilmark{1}
        Takao~\textsc{Soyano},\altaffilmark{10}
        Yoichi~\textsc{Tamura},\altaffilmark{1}
        Masuo~\textsc{Tanaka},\altaffilmark{1}
        Ken'ichi~\textsc{Tarusawa},\altaffilmark{10}
        and
        Yuzuru~\textsc{Yoshii}\altaffilmark{1}
       }
\altaffiltext{1}{Institute of Astronomy, The University of Tokyo, 2-21-1 Osawa, Mitaka, Tokyo 181-0015, Japan}
\email{konishi@ioa.s.u-tokyo.ac.jp}
\altaffiltext{2}{Astronomical Institute, Tohoku University, Aramaki, Aoba, Sendai 980-8578, Japan}
\altaffiltext{3}{Department of Astronomy, The University of Tokyo, 7-3-1 Hongo, Bunkyo-ku, Tokyo 113-0033, Japan}
\altaffiltext{4}{OptCraft, 3-16-8-101 Higashi Hashimoto, Midori-ku, Sagamihara, Kanagawa 252-0144, Japan}
\altaffiltext{5}{Department of Electrical Engineering and Center of Astro Engineering, Pontificia Universidad Cat$\acute{o}$lica de Chile, Av. Vicu$\tilde{n}$a Mackenna 4860, Macul, Santiago, Chile}
\altaffiltext{6}{National Astronomical Observatory of Japan, 2-21-1 Osawa, Mitaka, Tokyo 181-8588, Japan}
\altaffiltext{7}{Nagoya City Science Museum, 2-17-1 Sakae, Naka-ku, Nagoya 460-0008, Japan}
\altaffiltext{8}{Institute of Space and Astronautical Science, Japan Aerospace Exploration Agency, 3-1-1 Yoshinodai, Chuo, Sagamihara, Kanagawa 252-5210, Japan}
\altaffiltext{9}{Joint ALMA Observatory, Alonso de Cordova 3107, Vitacura, Santiago 763-0355, Chile}
\altaffiltext{10}{Kiso Observatory, Institute of Astronomy, School of Science, The University of Tokyo, 10762-30 Mitake, Kiso-machi, Kiso-gun, Nagano 397-0101, Japan}
\altaffiltext{11}{Department of Earth and Planetary Sciences, Faculty of Science, Kobe University, 1-1 Rokkodai-cho, Nada, Kobe 657-8501, Japan}
\altaffiltext{12}{Faculty of Science, Kagoshima University, 1-21-24 Korimoto, Kagoshima 890-8580, Japan}
\altaffiltext{13}{Center for Low Carbon Society Strategy, Japan Science and Technology Agency, 7 Goban-cho, Chiyoda-ku, Tokyo 102-0076, Japan}

\maketitle

\begin{abstract}
We have developed a near-infrared camera called ANIR (Atacama Near-InfraRed camera) for the University of Tokyo Atacama Observatory 1.0-m telescope (miniTAO) installed at the summit of Cerro Chajnantor (5,640 m above sea level) in northern Chile.
The camera provides a field of view of \timeform{5'.1} $\times$ \timeform{5'.1} with a spatial resolution of \timeform{0''.298} pixel$^{-1}$ in the wavelength range from 0.95 to 2.4 $\mu$m, using Offner relay optics and a PACE HAWAII-2 focal plane array.
Taking advantage of the dry site, the camera is capable of hydrogen Paschen-$\alpha$ (Pa$\alpha$, $\lambda=$ 1.8751 $\mu$m in air) narrow-band imaging observations, at which wavelength ground-based observations have been quite difficult due to deep atmospheric absorption mainly from water vapor.
We have been successfully obtaining Pa$\alpha$ images of Galactic objects and nearby galaxies since the first-light observation in 2009 with ANIR.
The throughputs at the narrow-band filters ($N$1875, $N$191) including the atmospheric absorption show larger dispersion ($\sim$ 10\%) than those at broad-band filters (a few percent), indicating that they are affected by temporal fluctuations in Precipitable Water Vapor (PWV) above the site.
We evaluate the PWV content via the atmospheric transmittance at the narrow-band filters, and derive the median and the dispersion of the distribution of the PWV of 0.40 $\pm$ 0.30 and 0.37 $\pm$ 0.21 mm for the $N$1875 and $N$191 data, respectively, which are remarkably smaller (49 $\pm$ 38\% for $N$1875 and 59 $\pm$ 26\% for $N$191) than radiometry measurements at the base of Cerro Chajnantor (an altitude of 5,100 m).
The decrease in PWV can be explained by the altitude of the site when we assume that the vertical distribution of the water vapor is approximated at an exponential profile with scale height within 0.3--1.9 km (previously observed values at night).
We thus conclude that miniTAO/ANIR at the summit of Cerro Chajnantor indeed provides us an excellent capability for a \textit{ground-based} Pa$\alpha$ observation.
\end{abstract}

\section{Introduction}
\label{sect:intro}

\subsection{Pa$\alpha$ emission as a tool for Unveiling the Dust-obscured Universe}
To explore the formation and the evolution of the present-day ``normal'' galaxies like our Galaxy, it is important to characterize where and how star formation occurs in a galaxy.
In the local universe, it is well known that star formation rate (SFR) correlates with dust opacity (\cite{Takeuchi10}; \cite{Bothwell11}), where intensely star-forming regions and galaxies are obscured by a huge amount of dust, and become optically thick (\cite{Alonso-Herrero06b}; \cite{Piqueras13}).
Therefore, the UV continuum emission and hydrogen recombination lines in the optical ranges such as H$\alpha$ at 0.6563 $\mu$m and H$\beta$ at 0.4861 $\mu$m emitted from those dusty star-forming regions are easily attenuated, and then it would be complicated to correctly understand the distribution of star-forming regions in the whole galaxy, especially when the galaxy has a patchy distribution of dust (\cite{Garcia06}).

On the other hand, Paschen-$\alpha$ (Pa$\alpha$) at 1.8571 $\mu$m (in air; 1.8756 $\mu$m in vacuum), the strongest emission line in the near-infrared (NIR, $\lambda$ $\sim$ 0.8--2.5 $\mu$m) wavelength range, is less affected by dust thanks to its longer wavelength than the optical lines (\cite{Kennicutt98}), and becomes a powerful and direct indicator of SFR, especially in dusty regions.
In particular, the observed Pa$\alpha$ emission becomes stronger than H$\alpha$ at $E(B-V) >$ 1.2 ($A_{V} >$ 3.7) and than Br$\gamma$ at $E(B-V) <$ 28.0 ($A_{V} <$ 86.2) with an assumption of a Milky-Way-like extinction curve and $A_{V}/E(B-V) = $ 3.08 (\cite{Pei92}), while the intrinsic Pa$\alpha$ luminosity is 0.12 of H$\alpha$, 2.14 of Pa$\beta$ (1.2818 $\mu$m in air), and 12.4 of Br$\gamma$ (2.1655 $\mu$m in air) for Case B recombination with an electron temperature of $10^{4}$ K and density of $10^{4}$ cm$^{-3}$ (\cite{Osterbrock89}).

\begin{figure}
  \begin{center}
    \FigureFile(80mm,80mm){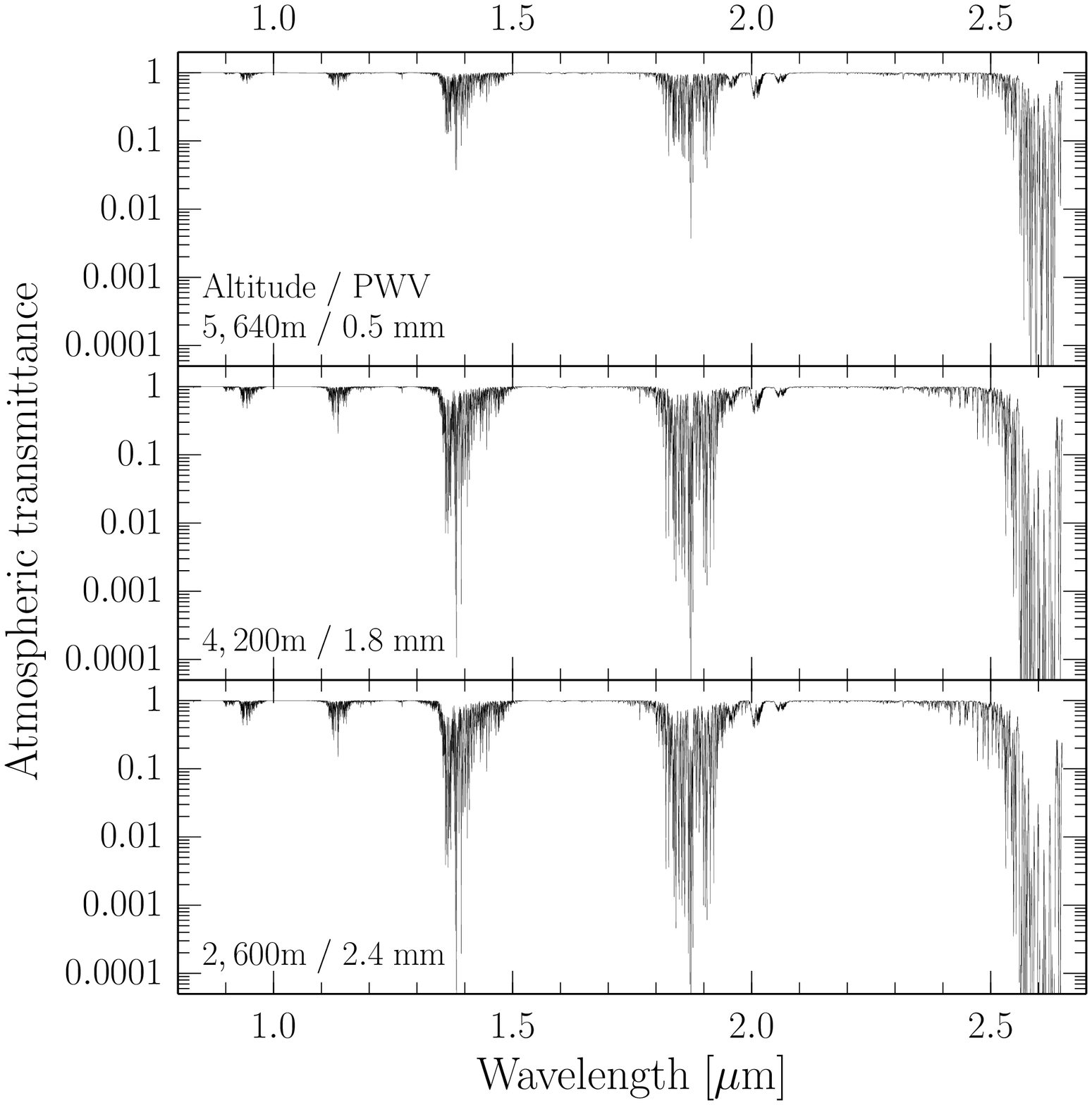}
  \end{center}
  \caption{Atmospheric transmittances in the NIR wavelength range at TAO (\textit{top}), Mauna Kea (\textit{middle}, 4,200 m above sea level) and Paranal (\textit{bottom}, 2,600 m), simulated using an atmospheric transmission model, ATRAN (\cite{Lord92}) with a spectral resolution of $\sim$ 50000. Median PWV of 0.5 (\cite{Giovanelli01b}), 1.8 (\cite{Otarola10}), and 2.4 mm (\cite{Kerber10}) are assumed for TAO, Mauna Kea, and Paranal, respectively. Remarkable improvements of the transmittance are seen at around 1.9 $\mu$m as well as at gaps between conventional broad atmospheric windows such as $J$ (1.28 $\mu$m), $H$ (1.67 $\mu$m), and $K_{\rm{s}}$-band (2.15 $\mu$m).}
  \label{fig:atran_nir}
\end{figure}

However, Pa$\alpha$ observations from the ground have been quite difficult so far due to deep atmospheric absorption features around their wavelength, mostly caused by water vapor, and hence most Pa$\alpha$ observational studies have been limited to those by Near Infrared Camera and Multi-Object Spectrometer (NICMOS, \cite{Thompson98}) on board the \textit{Hubble} Space Telescope (\textit{HST}).
For example, \citet{Alonso-Herrero06b} use the NICMOS Pa$\alpha$ narrow-band (F187N and F190N) imaging data to study star formation properties of dusty star-forming galaxies in the local universe with a high spatial resolution, and find the compact distributions of Pa$\alpha$ emission along various galaxy structures such as nucleus and spiral arm. They also establish a linear empirical relation between Pa$\alpha$ and 24 $\mu$m luminosity as well as the total infrared luminosity as an indicator of SFR for those galaxies.
In addition, a wide-field Pa$\alpha$ imaging survey of the Galactic Center (\cite{Wang10}; \cite{Dong11}) produces a high spatial-resolution map of stars and an ionized diffuse gas with a possible new class of massive stars not associated with any known star clusters.
Recently, a NIR camera and spectrograph, FLITECAM (\cite{Mclean12}), for NASA's Stratospheric Observatory for Infrared Astronomy (SOFIA) has been developed, and will support Pa$\alpha$ imaging observations with two narrow-band filters centered at 1.875 and 1.900 $\mu$m.

As described above, Pa$\alpha$ observations are quite useful for a variety of targets obscured by dust and can provide a new insight into those studies, yet the amount of observing time and facilities available for that purpose are still very limited. Therefore, we have developed a new instrument at an appropriate astronomical site for \textit{ground-based} Pa$\alpha$ observations.

\begin{table}
  \caption{Specification of the ANIR NIR Unit.}
  \label{tab:spec_nir}
  \begin{center}
    \begin{tabular}{ll}
\hline
\hline
\rule[-1ex]{0pt}{3.5ex} Wavelength & 0.95--2.4 $\mu$m \\
\rule[-1ex]{0pt}{3.5ex} Detector & PACE HAWAII-2 \\
\rule[-1ex]{0pt}{3.5ex} Pixel format & 1024 $\times$ 1024\footnotemark[$*$] \\
\rule[-1ex]{0pt}{3.5ex} Pixel pitch & 18.5 $\mu$m \\
\rule[-1ex]{0pt}{3.5ex} Field of view & \timeform{5'.1} $\times$ \timeform{5'.1} \\
\rule[-1ex]{0pt}{3.5ex} Pixel scale & \timeform{0''.298} pixel$^{-1}$ \\
\rule[-1ex]{0pt}{3.5ex} Broad-band filters\footnotemark[$\dagger$] & $Y$, $J$, $H$, $K_{\rm{s}}$ \\
\rule[-1ex]{0pt}{3.5ex} Narrow-band filters & $N$128, $N$1875, $N$191, $N207$ \\
\hline
\multicolumn{2}{@{}l@{}}{\hbox to 0pt{\parbox{80mm}{\footnotesize
\rule[-1ex]{0pt}{3.5ex} \footnotemark[$*$] The actual format size of HAWAII-2 FPA is 2048 $\times$ 2048 pixels, of which we use only 1024 $\times$ 1024 pixels.

\rule[-1ex]{0pt}{3.5ex} \footnotemark[$\dagger$] The $J$, $H$, and $K_{\rm{s}}$ filters are those of Mauna Kea Observatories (MKO) infrared filter set (\cite{Simons02}; \cite{Tokunaga02}).
}\hss}}
    \end{tabular}
  \end{center}
\end{table}

\subsection{ANIR: Ground-based Pa$\alpha$ Imager for the miniTAO 1.0-m Telescope}

\begin{figure*}[!ht]
  \begin{center}
    \FigureFile(170mm,170mm){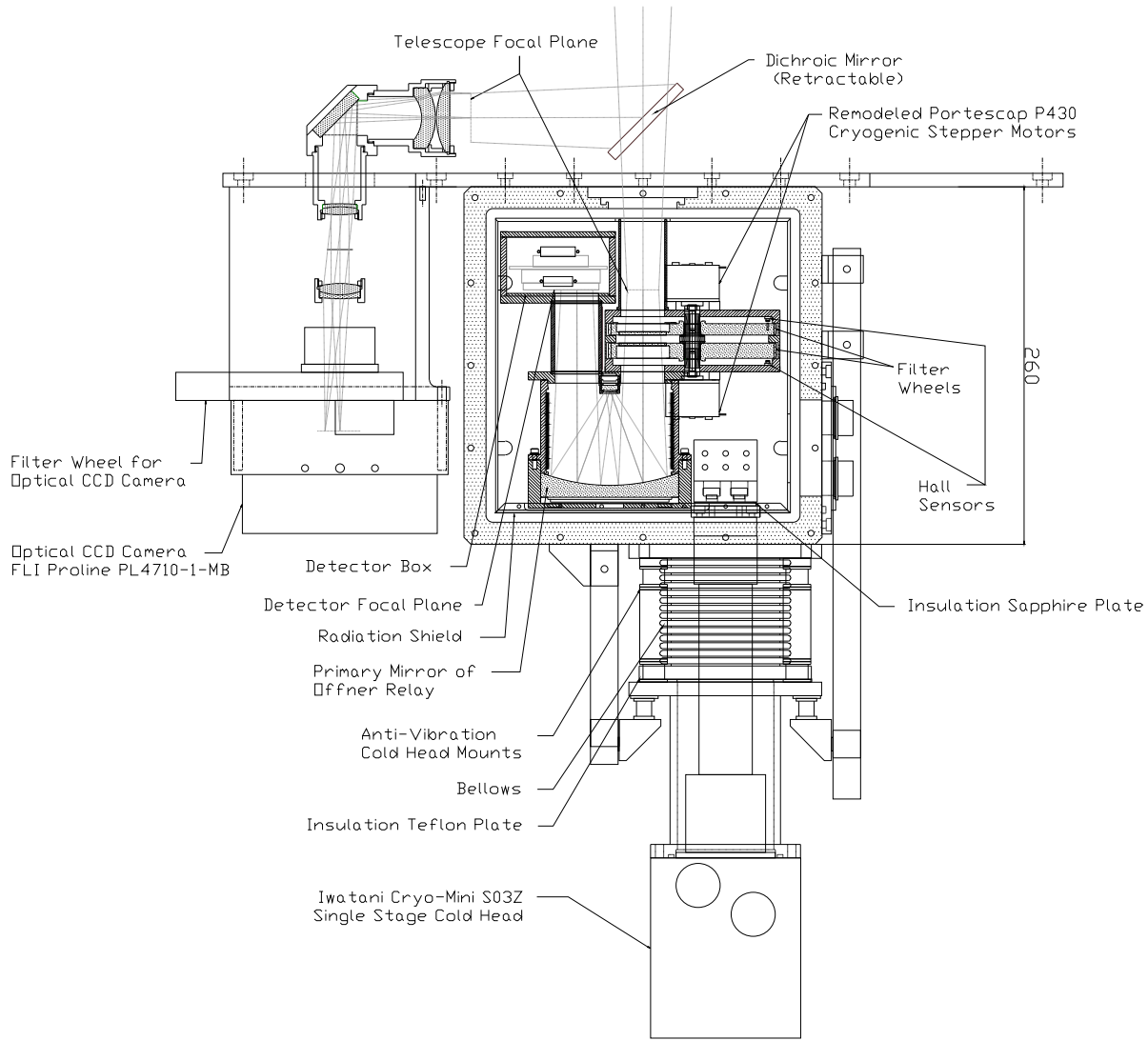}
  \end{center}
  \caption{Schematic drawing of ANIR.}
  \label{fig:anir_2d}
\end{figure*}

The University of Tokyo Atacama Observatory (TAO, \cite{Yoshii10}) 1.0-m telescope, called miniTAO, is an optical/infrared telescope installed in 2009, located at the summit of Cerro Chajnantor (an altitude of 5,640 m or 18,500 ft above sea level) in northern Chile as the world's highest astronomical observatory (\cite{Sako08a}; \cite{Minezaki10}).
The site was expected to be very dry from satellite data reported by \citet{Erasmus02} (precipitable water vapor, PWV, of 0.5 mm at 25th percentile) and from radiosonde measurements by \citet{Giovanelli01b} (0.5 mm at a median above an altitude of 5,750 m).
Figure \ref{fig:atran_nir} shows a comparison of simulated atmospheric transmittances between the TAO site and the other sites at lower altitudes, in which there are remarkable improvements of the transmittance at gaps between conventional broad atmospheric windows, especially around 1.9 $\mu$m, i.e., Pa$\alpha$ wavelength.

The Atacama NIR camera (ANIR)\footnote{ANIR web site: http://www.ioa.s.u-tokyo.ac.jp/kibans/anir\_en/} is one of the instruments for the Cassegrain focus of the miniTAO telescope.
It is capable of wide-field Pa$\alpha$ narrow-band (NB) imaging observations as well as ordinary broad-band imaging one.
It also has a capability of optical imaging and slitless spectroscopic observations simultaneously with the NIR imaging using a retractable dichroic mirror.
We have achieved first-light observations using NB filters targeted on the Pa$\alpha$ emission line (Figure \ref{fig:atran_paa}) in 2009, and now ANIR is in operation mainly for Pa$\alpha$ observations of the Galactic center/plane and nearby starburst galaxies in order to understand the star formation activities hidden in thick dust clouds (e.g., \cite{Tateuchi12b}; \cite{Komugi12}).

\begin{figure*}[!ht]
  \begin{center}
    \FigureFile(170mm,170mm){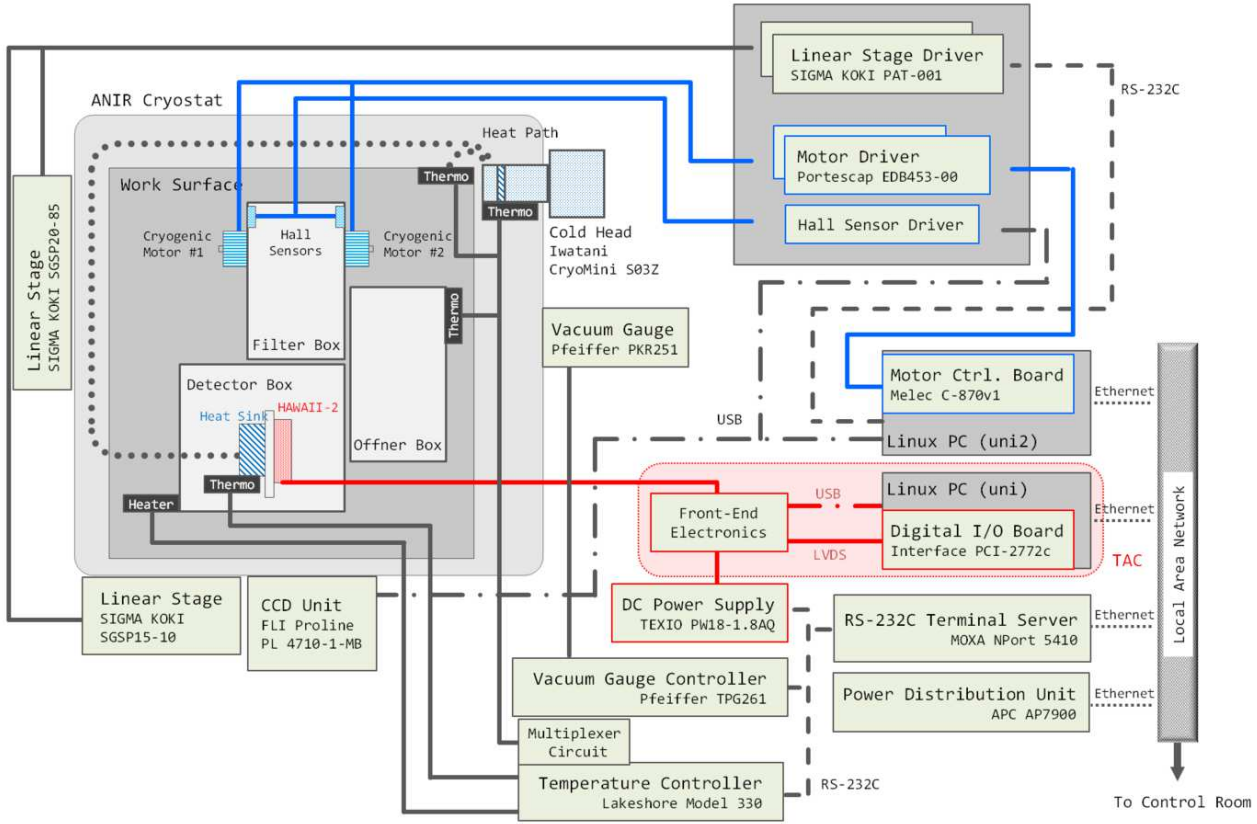}
  \end{center}
  \caption{Hardware block diagram of ANIR.}
  \label{fig:hardware_diagram}
\end{figure*}

In this paper, we describe the overall design of ANIR in Section \ref{sect:instrument}, its performance evaluated through laboratory tests and actual observations at the TAO site in Section \ref{sect:performance}, and a site evaluation through the Pa$\alpha$ imaging data in Section \ref{sect:pwv}.
We refer the reader to \citet{Tateuchi12a} and \citet{Tateuchi14} for a detailed description of data reduction processes and quantitative analysis of Pa$\alpha$ data.

\section{Instrument}
\label{sect:instrument}

\subsection{Near-infrared Imager Unit}
\label{sect:instrument_nir}

Here we describe the design of the NIR unit providing broad-band and NB imaging functions in the NIR.
A hardware design of the optical unit is described in Section \ref{sect:instrument_opt}.

\subsubsection{Overview}

Figure \ref{fig:anir_2d} shows a cross-sectional view of ANIR, and a block diagram of the hardware is shown in Figure \ref{fig:hardware_diagram}.
The specifications of the NIR unit are summarized in Table \ref{tab:spec_nir}.
The unit covers a field of view (FoV) of \timeform{5'.1} $\times$ \timeform{5'.1} with a spatial resolution of \timeform{0''.298} pixel$^{-1}$ using an engineering grade Producible Alternative to CdTe for Epitaxy (PACE) HAWAII-2 array detector, which is a 2048 $\times$ 2048 pixel HgCdTe NIR focal plane array (FPA) manufactured by Teledyne Scientific \& Imaging LLC.
Note that only a single quadrant with 1024 $\times$ 1024 pixels is used.
The filter set consists of four standard broad-band filters ($Y$, $J$, $H$, and $K_{\rm{s}}$) and four NB filters ($N$128, $N$1875, $N$191, and $N$207) as listed in Table \ref{tab:spec_nir_NB}.
The $N$1875 filter has a central wavelength of $\lambda_{\rm{c}}=$ 1.8759 $\mu$m and a bandwidth of $\Delta\lambda=$ 0.0079 $\mu$m, which covers the Pa$\alpha$ ($\lambda=$1.8751 $\mu$m) with radial velocities within $\sim$ $-$580 to $+$680 km s$^{-1}$.
On the other hand, the $N$191 filter ($\lambda_{\rm{c}}$ = 1.911 $\mu$m, $\Delta\lambda=$ 0.033 $\mu$m) corresponds to redshifted Pa$\alpha$ lines with recession velocities of c$z$ $\sim$ 2900--8200 km s$^{-1}$.
The $N$191 filter is also used for taking off-band (continuum) data for the $N$1875 data, and vice versa.

\begin{figure*}
  \begin{center}
    \FigureFile(85mm,85mm){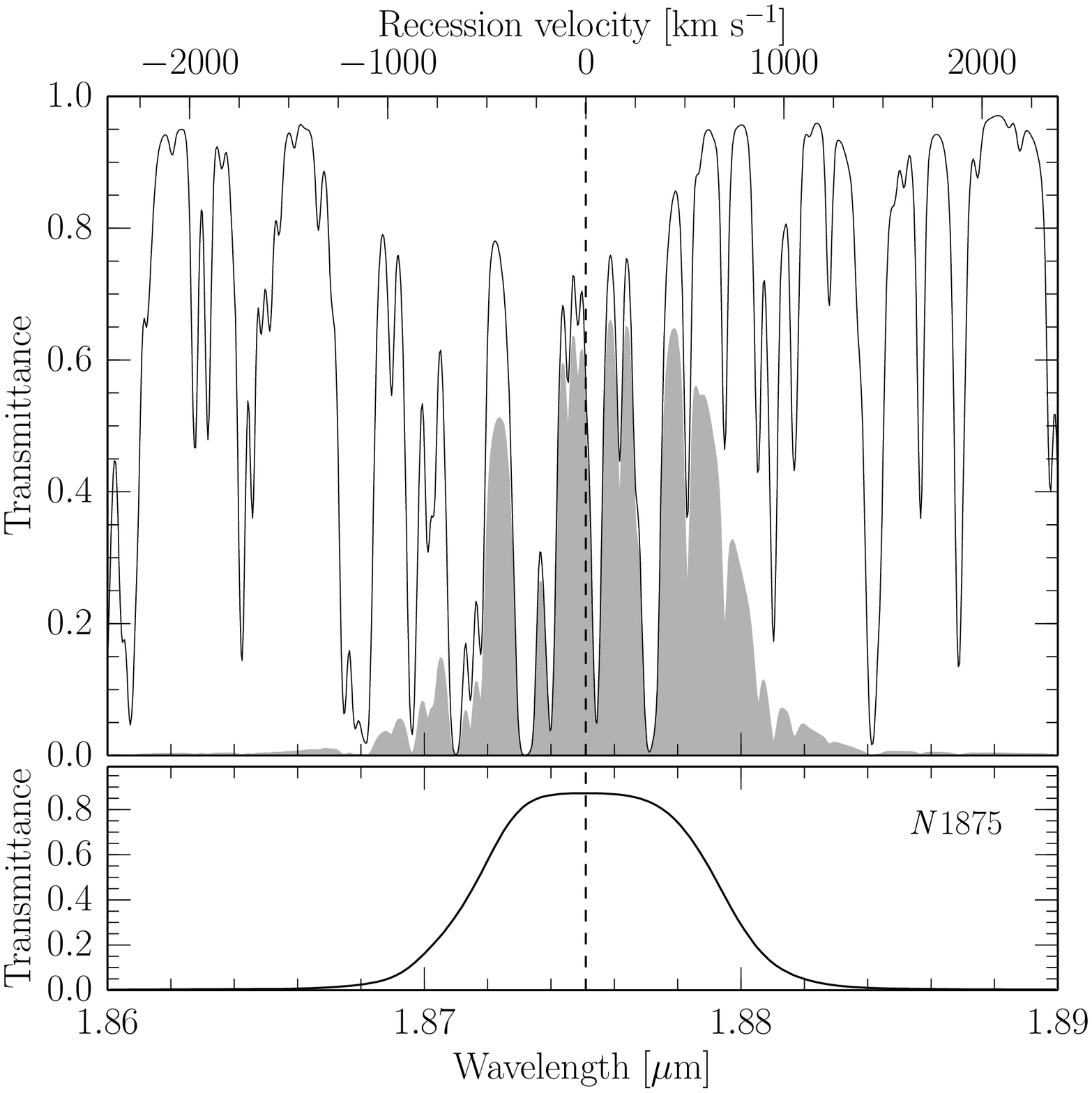}
    \FigureFile(85mm,85mm){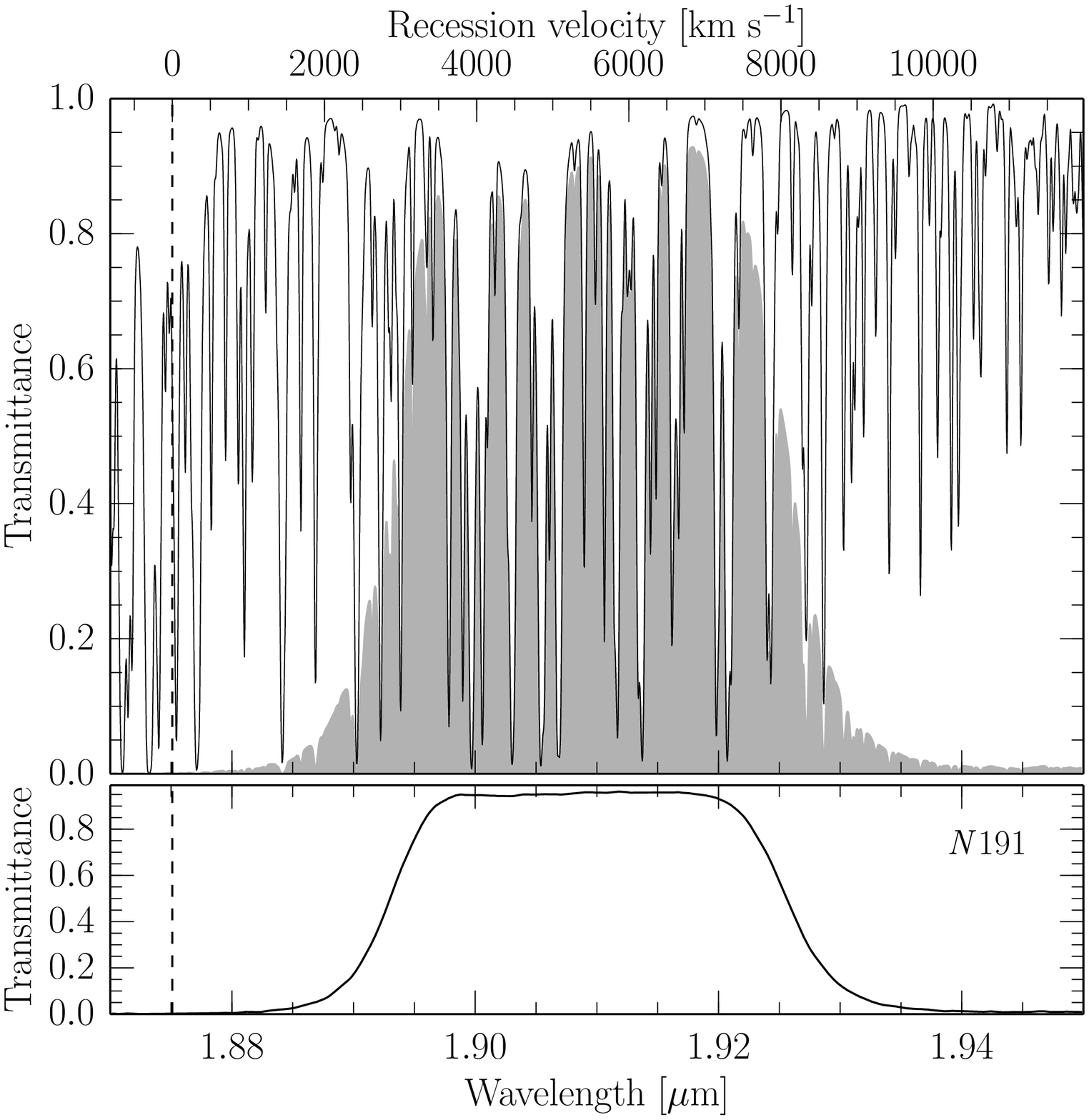}
  \end{center}
  \caption{Atmospheric transmittances at the TAO site simulated with the PWV of 0.5 mm are shown for the $N$1875 (\textit{left}) and $N$191 (\textit{right}) filters, respectively. The shaded region in the \textit{top} panels represents the effective transmittance including the atmospheric transmittance (solid line) and the filter transmittance measured at 77 K (shown in the \textit{bottom} panel). The vertical dashed line indicates the position of Pa$\alpha$ wavelength (1.8751 $\mu$m). The average atmospheric transmittance within the bandpass ($\Delta \lambda$) around the center ($\lambda_{\rm{c}}$) listed in Table \ref{tab:spec_nir_NB} is 0.48 and 0.64 at the $N$1875 and $N$191 filters, respectively.}
  \label{fig:atran_paa}
\end{figure*}

Figure \ref{fig:atran_paa} shows details of the atmospheric transmittances shown in the top panel of Figure \ref{fig:atran_nir} at the wavelength range of interest for the Pa$\alpha$ NB filters.
While there are absorption features mostly by water vapor even at the altitude of 5,640 m, several windows with a high transmittance exist.
The average atmospheric transmittances at the PWV of 0.5 mm are approximately 0.48 and 0.64 within the bandpasses of the $N$1875 and $N$191 filters, respectively.

\begin{table*}
  \caption{NIR narrow-band filters.}
  \label{tab:spec_nir_NB}
  \begin{center}
    \begin{tabular}{lccccl}
\hline
\hline
\rule[-1ex]{0pt}{3.5ex} Name & \multicolumn{4}{c}{Wavelength [$\mu$m]} & Targeted Line \\
\rule[-1ex]{0pt}{3.5ex}  & \multicolumn{2}{c}{Specification} & \multicolumn{2}{c}{Measurement\footnotemark[$*$]} & \\
\rule[-1ex]{0pt}{3.5ex}  & Center ($\lambda_{\rm{c}}$\footnotemark[$\dagger$]) & Width ($\Delta\lambda$\footnotemark[$\ddagger$]) & Center & Width \\
\hline
\rule[-1ex]{0pt}{3.5ex} $N$128 & 1.2818 & 0.0210 & 1.2814 & 0.0217 & Pa$\beta$ (1.2818 $\mu$m) \\
\rule[-1ex]{0pt}{3.5ex} $N$1875 & 1.8751 & 0.0080 & 1.8759 & 0.0079 & Pa$\alpha$ (1.8751 $\mu$m)\\
\rule[-1ex]{0pt}{3.5ex} $N$191 & 1.9010 & 0.0306 & 1.9105 & 0.0329 & Pa$\alpha$ off-band\footnotemark[$\S$] \\
\rule[-1ex]{0pt}{3.5ex} $N$207 & 2.0750 & 0.0400 & 2.0742 & 0.0391 & C\emissiontype{IV} (2.078 $\mu$m)\\
\hline
\multicolumn{4}{@{}l@{}}{\hbox to 0pt{\parbox{115mm}{\footnotesize
\rule[-1ex]{0pt}{3.5ex} \footnotemark[$*$] Measured at 77 K.

\rule[-1ex]{0pt}{3.5ex} \footnotemark[$\dagger$] Isophotal wavelength (\cite{Tokunaga05}).

\rule[-1ex]{0pt}{3.5ex} \footnotemark[$\ddagger$] Full-width at half maximum.

\rule[-1ex]{0pt}{3.5ex} \footnotemark[$\S$] Also corresponding to Pa$\alpha$ rest-frame wavelength emitted from extra-galactic objects within c$z\sim$ 2900--8200 km s$^{-1}$.
}\hss}}
    \end{tabular}
  \end{center}
\end{table*}

\subsubsection{Cryogenics}
\label{sect:cryogenics}

The cryostat is a compact cube with approximately 260 mm on a side, in which Offner relay optics (Section \ref{sect:optics}), a filter box with two wheels, and the HAWAII-2 FPA are housed, as shown in Figure \ref{fig:anir_2d}.
All the components in the cryostat are cooled down to 70 K to reduce thermal radiation, especially for observations in the $K_{\rm{s}}$-band.
A single-stage closed-cycle mechanical cooler with a cooling capacity of 25 W at 77 K is equipped.
A cold head anti-vibration mount and bellows are inserted between the cold head and the cryostat to reduce vibration caused by the cold head.
In addition, a Teflon plate and a sapphire plate are inserted to electrically insulate the cryostat from the cold head.
It takes approximately 24 hours to cool down and stabilize the cryostat ($\Delta T \lesssim 0.05$ K at the detector box) from the ambient temperature down to 70 K.

The two filter wheels are installed just after the focal plane of the telescope.
Each filter wheel has five slots with $\phi$ 34 mm.
A stepping motor controls its rotation, and neodymium magnets and a hall-effect sensor sense the position of the slots.

\subsubsection{Optics}
\label{sect:optics}

\begin{figure*}[!t]
  \begin{center}
    \FigureFile(85mm,85mm){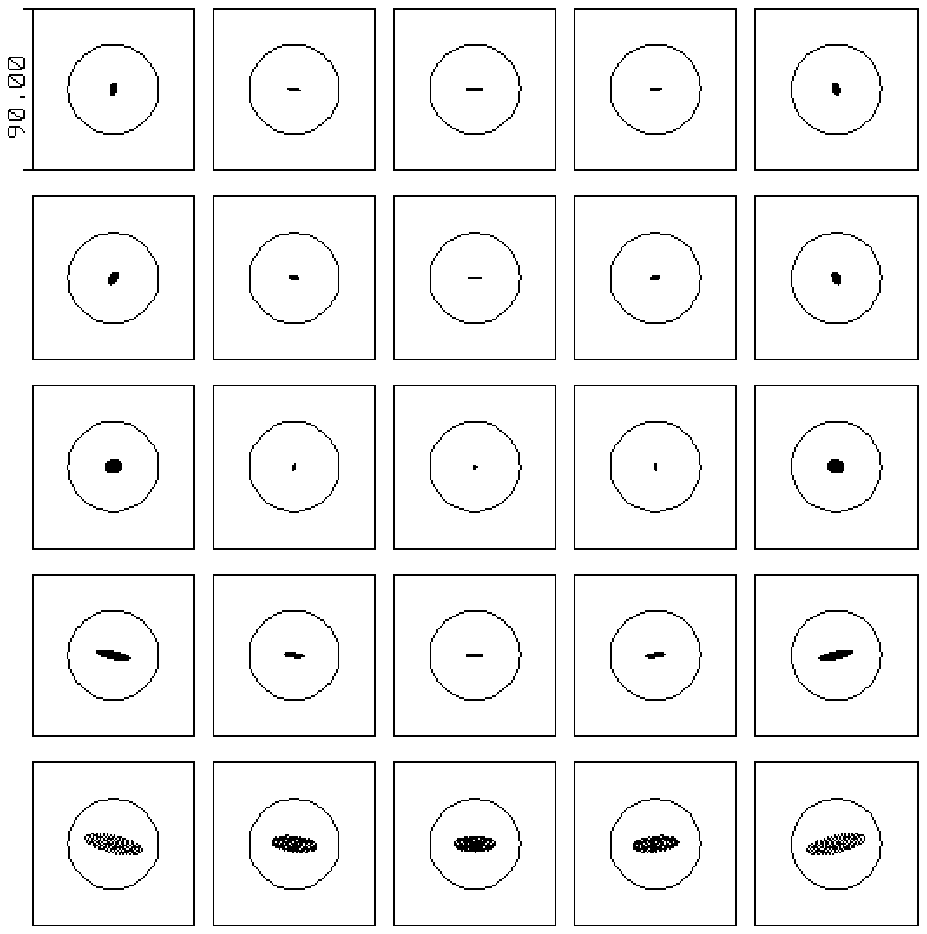}
    \FigureFile(85mm,85mm){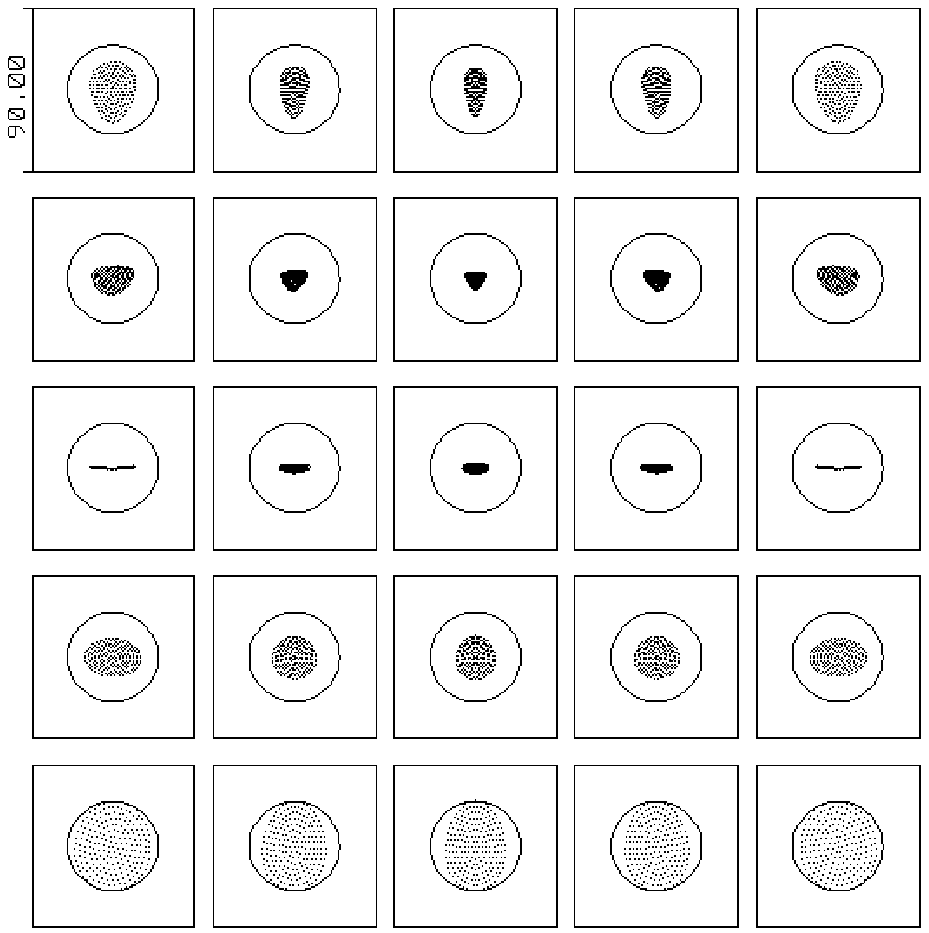}
  \end{center}
  \caption{Spot diagrams at $\lambda$ = 1.65 $\mu$m without (\textit{left}) and with (\textit{right}) the dichroic mirror. Diagrams at the center of the FoV and positions at a 5 $\times$ 5 grid with a step of $\pm$\timeform{1'.32} from the center are shown. A side of each box corresponds to 90 $\mu$m or \timeform{1''.5}. The circle corresponds to the Airy disk at $\lambda$ = 1.65 $\mu$m. It is worth mentioning that no significant image degradation caused by the insertion of the dichroic mirror is seen throughout wavelengths between 0.95 and 2.4 $\mu$m.}
  \label{fig:nir_sd}
\end{figure*}

Reflective Offner relay optics are employed for re-imaging, consisting of two (concave primary and convex secondary) spherical mirrors.
The primary mirror forms an image of the telescope pupil on the secondary mirror, which works as a cold Lyot stop.
The specifications of the mirrors are summarized in Table \ref{tab:offnerspec}.
These mirrors are gold-coated to achieve high reflectivity.

\begin{figure}
  \begin{center}
    \FigureFile(80mm,80mm){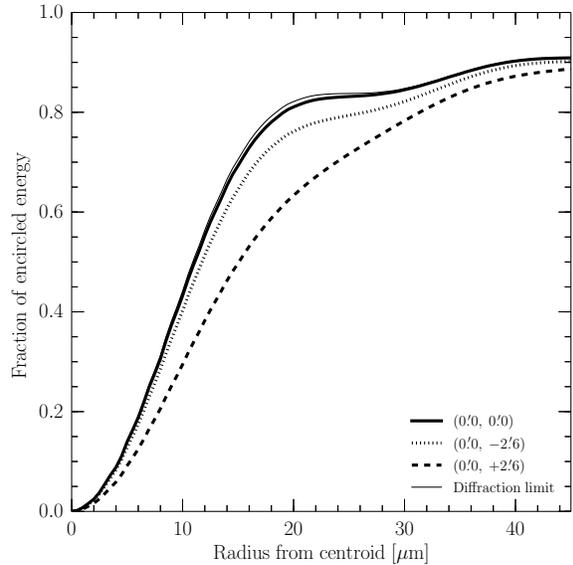}
  \end{center}
  \caption{Encircled energy distributions of the point spread functions of the NIR unit at the positions (\timeform{0'.0}, \timeform{-2'.6}) and (\timeform{0'.0}, \timeform{+2'.6}) from the center of the FoV (\timeform{0'.0}, \timeform{0'.0}) (thick lines) in the case that the dichroic mirror is inserted, compared to that of a diffraction-limited image (thin solid line).}
  \label{fig:nir_ece_wDM}
\end{figure}

Figure \ref{fig:nir_sd} shows spot diagrams at $\lambda$ = 1.65 $\mu$m across the FoV.
Sharp image quality is achieved at any positions within the FoV.
As shown in the right-hand panel of Figure \ref{fig:nir_sd}, even with a dichroic mirror inserted for simultaneous optical-NIR observations (Section \ref{sect:instrument_opt}), spot sizes are still smaller than or comparable to the size of the Airy disk.
Figure \ref{fig:nir_ece_wDM} shows encircled energy distributions, the fraction of energy in a given radius to the total energy from a point source, for the center and off-center spots with the dichroic mirror inserted, compared to that of a diffraction-limited spot.
We confirm that the diameter encircling the 80\% energy is less than 1.7 pixels ($\sim$ 31.5 $\mu$m) or \timeform{0''.5} across the FoV.

\begin{table}[t]
  \caption{Specification of the Offner relay optics.}
  \label{tab:offnerspec}
  \begin{center}
    \begin{tabular}{llr}
\hline
\hline
\rule[-1ex]{0pt}{3.5ex} Primary mirror & Radius of curvature & 140 mm \\
\rule[-1ex]{0pt}{3.5ex}  & Effective diameter & 90 mm \\
\rule[-1ex]{0pt}{3.5ex} Secondary mirror & Radius of curvature & 70 mm \\
\rule[-1ex]{0pt}{3.5ex}  & Effective diameter & 9 mm \\
\rule[-1ex]{0pt}{3.5ex} Offset of optical axis & & 24 mm \\
\hline
    \end{tabular}
  \end{center}
\end{table}

\subsubsection{Data Acquisition and Control System}

\begin{figure*}
  \begin{center}
    \FigureFile(170mm,170mm){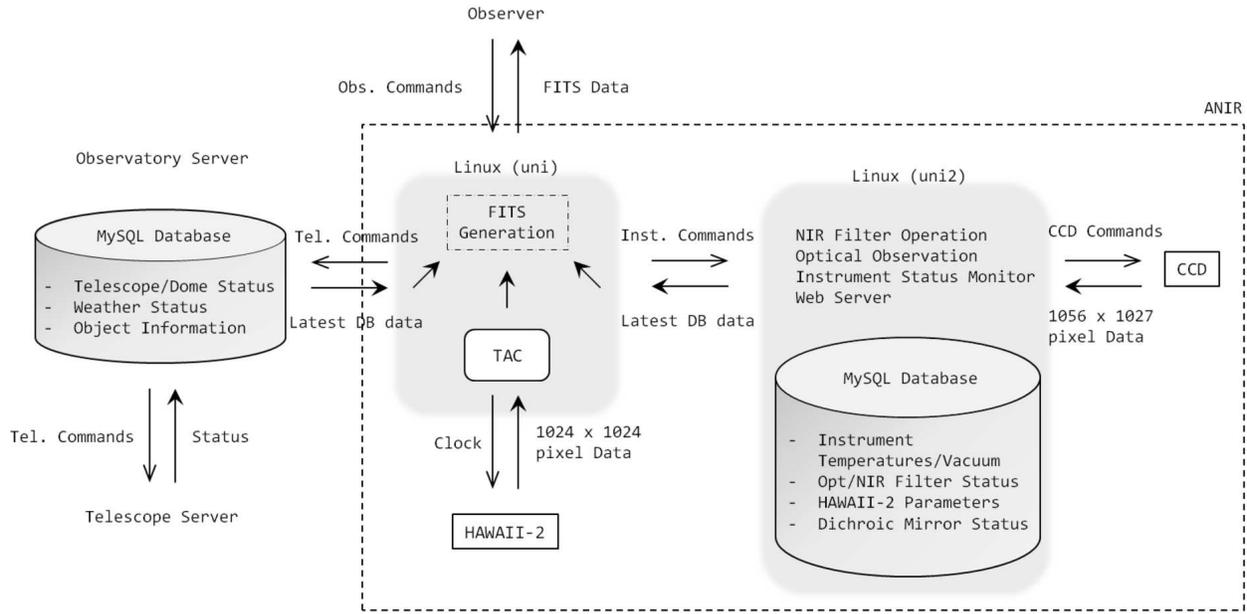}
  \end{center}
  \caption{Software block diagram of ANIR.}
  \label{fig:software_diagram}
\end{figure*}

ANIR is operated by two Linux PCs (named \textit{uni} and \textit{uni2}); one is dedicated to control HAWAII-2 FPA (TAO Array Controller, TAC), and the other handles all the other tasks, including operation of the filter wheels and the optical unit (see Section \ref{sect:instrument_opt}), acquisition of house keeping information such as temperature and vacuum pressure, and management of various instrument status.
The TAC system (see \cite{Sako08b} for more details) is a high-speed and flexible array controller using a real-time operating system instead of conventional dedicated processor devices such as digital signal processors (DSPs).
One core of a dual-core CPU in \textit{uni} is assigned to the real-time data processing, and engages in clock pattern generation and processing of acquired frame data.
This allows us to control the FPA in real time without affecting the performance of the operating system and other software.
A rate to read a pixel (Pixel rate) can be set to 3 to 8 $\mu$s.
Considering increasing readout noise with faster readout (slower Pixel rate), the rate of 4 $\mu$s is usually used, corresponding to a readout time of 1024 $\times$ 1024 pixels of 4.2 s, which is the shortest exposure time of the NIR unit.

Figure \ref{fig:software_diagram} summarizes the functions and gives a command/data flow diagram of the control system.
To efficiently handle status data increasing with time, we adopt a relational database management system, MySQL.
An successful example of the application of MySQL in astronomical instrumentation is described in \citet{Yoshikawa06}.
Several database tables are prepared in MySQL, and information concerning instrument status (filters used, insertion of the dichroic mirror, temperatures/pressures) is stored into the respective tables every minute.
When the exposure command is executed, a FITS (Flexible Image Transport System) header is constructed by collecting the latest information from those tables.

\subsection{Optical Unit}
\label{sect:instrument_opt}

To benefit from the advantages of the site, such as good seeing condition with \timeform{0.''7} at $V$-band at a median (\cite{Motohara08}; see also \cite{Giovanelli01a}), ANIR is capable of optical and NIR simultaneous imaging observations by inserting a dichroic mirror in front of the entrance window of the cryostat of the NIR unit (the upper part of Figure \ref{fig:anir_2d}).

\begin{figure*}[!t]
  \begin{center}
    \FigureFile(85mm,85mm){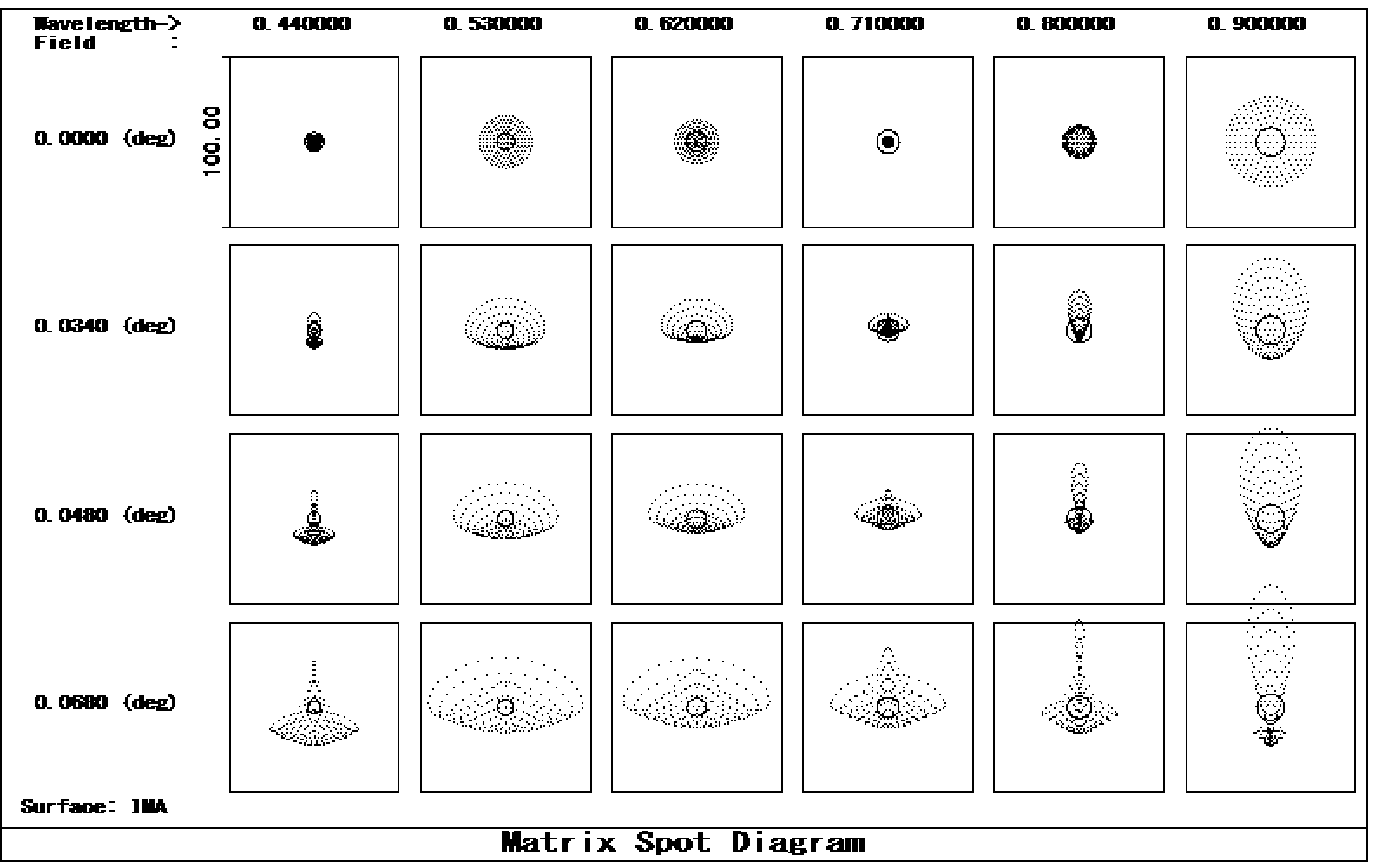}
    \FigureFile(85mm,85mm){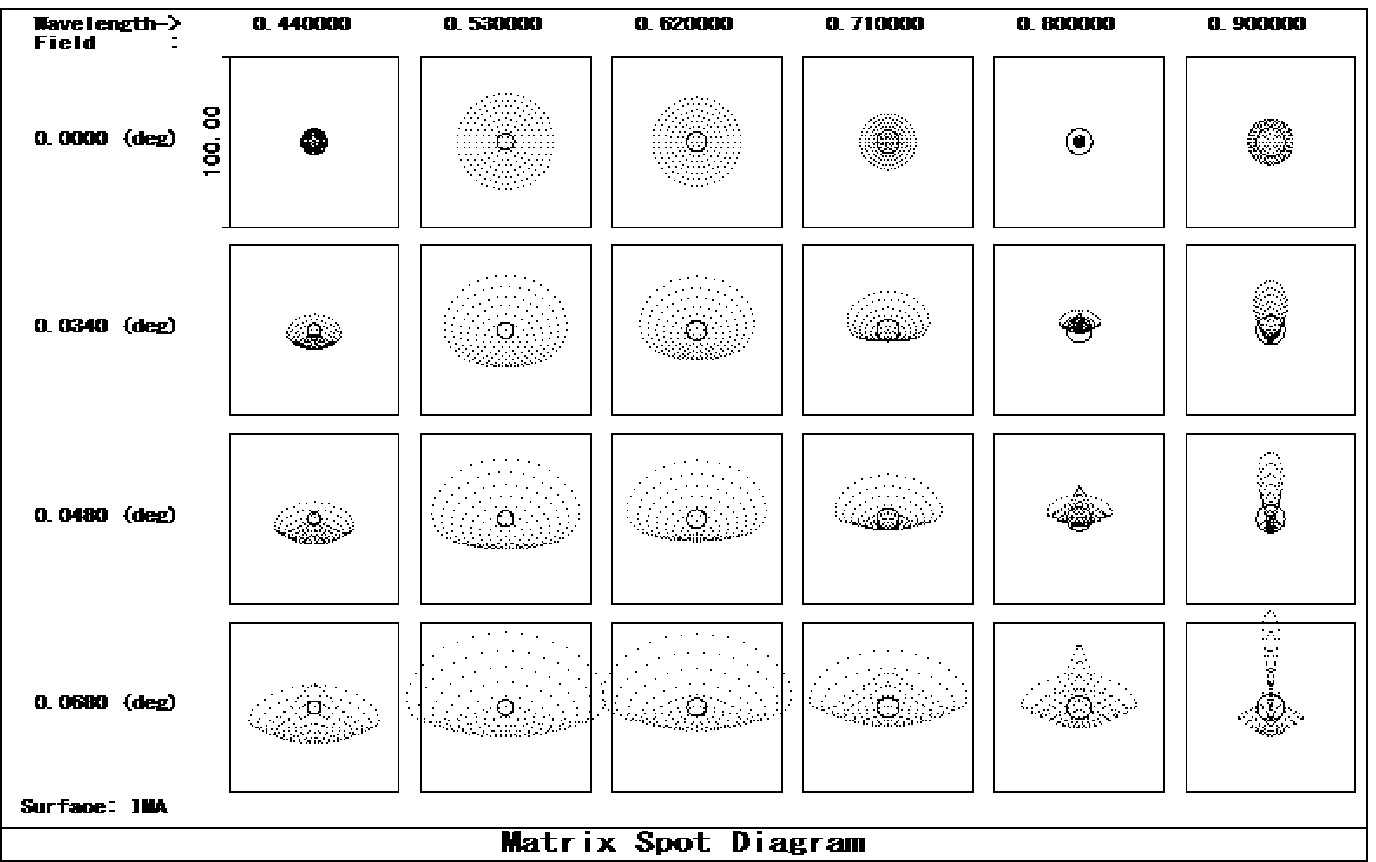}
  \end{center}
  \caption{\textit{Left}: Spot diagrams of the optical unit at the focus position. For each panel, a diagram for six wavelengths (0.44, 0.53, 0.62, 0.71, 0.80, and 0.90 $\mu$m) and four field positions (the center, and \timeform{2'.04}, \timeform{2'.88}, \timeform{4'.08} from the center) is shown. A side of each box corresponds to 100 $\mu$m or \timeform{2''.5}. \textit{Right}: Same as the \textit{left-hand} panel, but with the focus position offset by 0.2 mm to improve the performance in the $I$-band (0.81 $\mu$m).}
  \label{fig:opt_sd}
\end{figure*}

The specifications of the optical unit are summarized in Table \ref{tab:spec_opt}.
The unit covers a FoV of \timeform{6'.0} $\times$ \timeform{5'.9} with a spatial resolution of \timeform{0''.343} pixel$^{-1}$ using a commercial, peltier-cooling CCD camera unit incorporating a E2V backside-illuminated CCD, Proline PL4710-1-MB, fabricated by Finger Lakes Instrumentation.
Four Johnson-system broad-band filters ($B$, $V$, $R$, and $I$) for imaging and a low-resolution transmission grating (grism) with 75 lines mm$^{-1}$ and a blaze angle of 4.3$^{\circ}$ for slitless spectroscopy are available.
The dichroic mirror has a dimension of 50 $\times$ 71 $\times$ 10 mm with a wedge angle of 0.78$^\circ$ to minimize astigmatism in the NIR image.
Its reflectance at 0.4--0.86 $\mu$m is higher than 0.9 with an incident angle of 45$^{\circ}$, while its transmittance at 0.95--2.4 $\mu$m is kept higher than 0.9.
Figure \ref{fig:opt_sd} shows spot diagrams of the optical unit at the final focal plane.
Due to chromatic aberration of the re-imaging refractive optics, the focal position is offset by $\sim$ 0.2 mm in the $I$-band, which can be effectively compensated as shown in the right-hand panel of Figure \ref{fig:opt_sd} by shifting the camera optics mounted on a linear stage.
The dichroic mirror is mounted on a linear stage, and can be retracted to derive better-quality (higher efficiency, lower thermal background) NIR data when no simultaneous optical observation is required.

\section{Performance}
\label{sect:performance}

In this section, we describe the imaging performance of the NIR and optical units obtained through observations carried out in 2009--2011.

\subsection{Detector Performances}

We measure the conversion factor of the HAWAII-2 FPA readout system using flat frames with increasing exposure time in the laboratory, and obtain 3.2 $e^{-}$ ADU$^{-1}$ from the calculated mean and dispersion of pixel values.
The linearity of the FPA is evaluated from other flat images taken by increasing exposure time.
We ensure that the linearity is kept within $\sim$ 1\% up to 18,000 ADU or 57,600 $e^{-}$.
The readout noise is measured on the telescope in dark frames with the shortest exposure (i.e., 4.2 s) which are images with the $J$-band and $N$1875 filter inserted, where no photons fall on the FPA because those bandpasses do not overlap each other.
We obtain 16.6 $e^{-}$ as the rms of their median values with the Correlated Double Sampling (CDS) readout method.
A multiple readout (``up-the-ramp'' sampling) method is also examined, and we obtain 13.9 $e^{-}$ with 16 readouts.
The dark current derived from dark frames with 300--500 s exposures is 0.28 $e^{-}$ s$^{-1}$ pixel$^{-1}$.
We should mention that the dark current of HAWAII and HAWAII-2 FPAs have \textit{persistence} and that the level of the \textit{persistence} tends to be dependent on incident flux (\cite{Hodapp96}; \cite{Finger98}; \cite{Motohara02}).
For example, the \textit{persistence} of a dark current pattern would appear strongly on frames with a NB filter taken after exposures of high background, such as ones with a broad-band filter.
The TAC system continues to reset the FPA with an interval of $\sim$ 0.1 s in order to reduce the \textit{persistence} effect even if no exposure command is issued.

\begin{table}[!b]
  \caption{Specification of the ANIR Optical Unit.}
  \label{tab:spec_opt}
  \begin{center}
    \begin{tabular}{ll}
\hline
\hline
\rule[-1ex]{0pt}{3.5ex} Wavelength & 4000--8600 \AA \\
\rule[-1ex]{0pt}{3.5ex} CCD Unit & FLI Proline PL4710-1-MB \\
\rule[-1ex]{0pt}{3.5ex}  & (E2V backside-illuminated CCD) \\
\rule[-1ex]{0pt}{3.5ex} Pixel format & 1056 $\times$ 1027 \\
\rule[-1ex]{0pt}{3.5ex} Pixel pitch & 13.5 $\mu$m \\
\rule[-1ex]{0pt}{3.5ex} Field of view & \timeform{6'.0} $\times$ \timeform{5'.9} \\
\rule[-1ex]{0pt}{3.5ex} Pixel scale & \timeform{0''.343} pixel$^{-1}$ \\
\rule[-1ex]{0pt}{3.5ex} Filters & $B$, $V$, $R$, $I$ (Johnson System) \\
\rule[-1ex]{0pt}{3.5ex} Grism & 75 lines mm$^{-1}$, blaze angle $=$ 4.3$^{\circ}$ \\
\hline
    \end{tabular}
  \end{center}
\end{table}

The performance of the optical CCD are also evaluated in a similar manner to that described above.
We take flat images with increasing exposure time in the laboratory, and obtain the conversion factor of 1.818 $e^{-}$ ADU$^{-1}$ from the calculated mean and dispersion of pixel values.
The readout noise is measured from bias frames to be 14.5 $e^{-}$ rms.
We obtain the dark current of 0.011 $e^{-}$ s$^{-1}$ pixel$^{-1}$ at the CCD temperature of $\sim$ 220 K.

\subsection{Imaging Quality}

We evaluate the image quality of the NIR unit using bright (but unsaturated) and isolated field stars in images taken under good seeing condition.
By matching positions on the image ($X_{i}$, $Y_{i}$) of about 20 point sources located over the FoV with coordinates (\textit{$\alpha$}$_{i}$, \textit{$\delta$}$_{i}$) obtained from the Two Micron All Sky Survey (2MASS) Point Source Catalog (PSC, \cite{Skrutskie06}), we obtain a pixel scale of \timeform{0''.298} pixel$^{-1}$ with a fitting uncertainty of $\sim$ \timeform{0.''1} pixel$^{-1}$.
Image distortion is confirmed to be negligible ($<$ 1 pixel) over the FoV, as designed.
We derive a typical FWHM size of 2.41 $\pm$ 0.24 pixels or \timeform{0''.72} $\pm$ \timeform{0''.07} in the $K_{\rm{s}}$-band where ellipticities of the stars are small and uniform (0.08 $\pm$ 0.04).

The image quality of the optical unit is evaluated in a similar manner.
Raw images have non-negligible image distortion, especially at their corners.
We evaluate and correct for the distortion in the similar manner for the NIR unit as described above, by matching the positions of stars with those in the \textit{HST} Guide Star Catalog (GSC, \cite{Lasker08}) version 2.3 and fitting them to the latter using fifth-order polynomials to take into account the image distortion.
The mapping uncertainties are $\sim$ 0.3 pixels.
We obtain a pixel scale of \timeform{0''.343} pixel$^{-1}$ at the center of the FoV for all the broad-band filters ($B$, $V$, $R$, and $I$), and a typical FWHM size of 3.04 $\pm$ 0.22 pixels or \timeform{1''.04} $\pm$ \timeform{0''.08} with a homogeneous ellipticity distribution (0.04 $\pm$ 0.08) in the $V$-band.

\begin{figure*}[!ht]
  \begin{center}
    \FigureFile(85mm,85mm){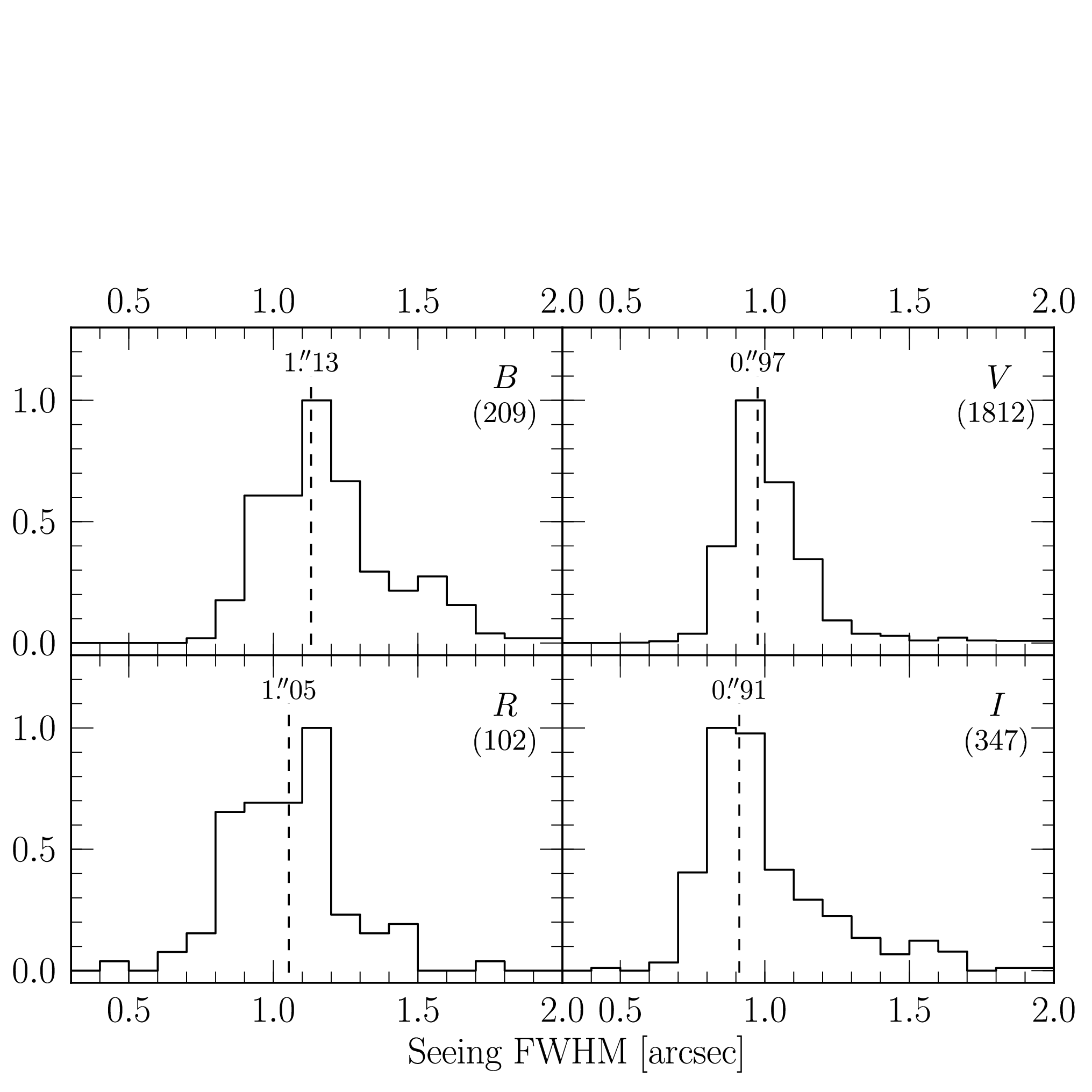}
    \FigureFile(85mm,85mm){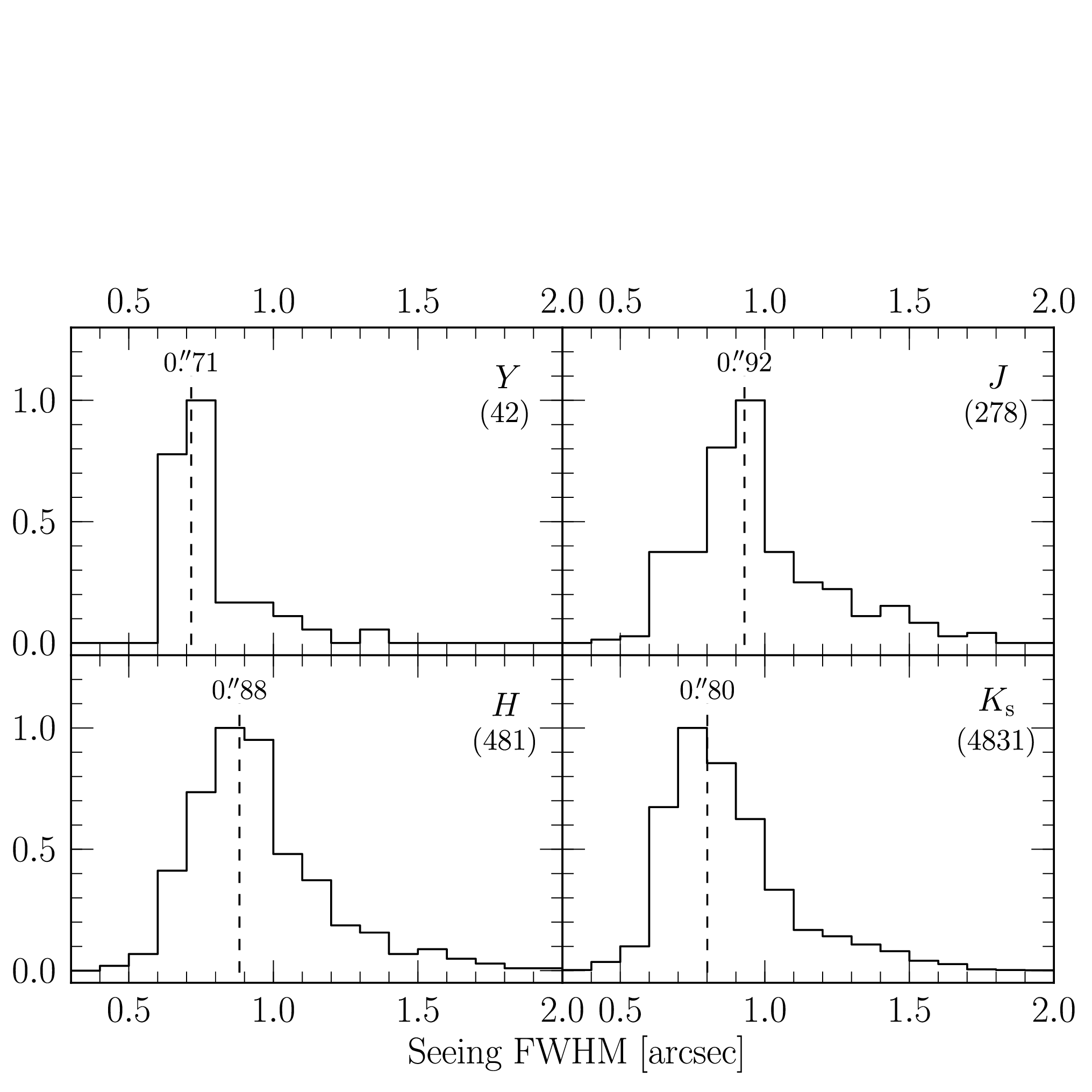}
  \end{center}
  \caption{Seeing statistics in the optical (\textit{left}) and NIR (\textit{right}) for miniTAO/ANIR. The histograms are normalized at their peak, and their median value is represented as the vertical dashed line. A number below the respective filter name indicates the number of frames used for the statistics.}
  \label{fig:seeing}
\end{figure*}

Figure \ref{fig:seeing} shows the seeing distribution for images taken in the latter half of the observation in 2010 and the former half in 2011.
The seeing size (FWHM) is measured using the IRAF \texttt{psfmeasure} task.
The median seeing size of $\sim$ \timeform{1''} or less is obtained for all the bands.
When we consider the diffraction-limited size and Hartmann constant (\timeform{0''.19}) of the telescope optics (\cite{Minezaki10}) and suppose that some data may be undersampled (FWHM $\lesssim$ 2 pixel) due to good seeing condition, the actual seeing distributions at the TAO site are expected to have a peak at a smaller size.

\subsection{Throughputs and Limiting Magnitudes}
\label{sect:throughput}

\begin{table*}
  \caption{Throughputs and limiting magnitudes\footnotemark[$*$].}
  \label{tab:performance}
  \begin{center}
    \begin{tabular}{lccccr}
\hline
\hline
\rule[-1ex]{0pt}{3.5ex}  & Throughput & \multicolumn{2}{c}{Limiting magnitude [mag]\footnotemark[$\dagger$]} & Sky brightness & $T_{\rm{BLIP}}$\footnotemark[$\ddagger$] \\
\rule[-1ex]{0pt}{3.5ex}  & [\%] & @ 60 s & @ 600 s & [mag arcsec$^{-2}$] & \multicolumn{1}{c}{[s]} \\
\hline
\rule[-1ex]{0pt}{3.5ex} $B$ & 17.5 & 21.3 & 23.5 & 22.0 & 2600 \\
\rule[-1ex]{0pt}{3.5ex} $V$ & 29.2 & 21.4 & 23.2 & 20.8 & 800 \\
\rule[-1ex]{0pt}{3.5ex} $R$ & 29.4 & 21.5 & 23.2 & 20.3 & 450 \\
\rule[-1ex]{0pt}{3.5ex} $I$ & 20.1 & 20.6 & 22.0 & 18.1 & 70 \\
\hline
\rule[-1ex]{0pt}{3.5ex} $Y$ & 14.3 & 18.8 & 20.1 & 16.8 & 130 \\
\rule[-1ex]{0pt}{3.5ex} $J$ & 18.9 & 19.5 & 20.9 & 16.4 & 50 \\
\rule[-1ex]{0pt}{3.5ex} $H$ & 29.1 & 18.4 & 19.7 & 14.6 & 10 \\
\rule[-1ex]{0pt}{3.5ex} $K_{\rm{s}}$ & 30.3 & 18.7 & 20.0 & 15.3 & 20 \\
\rule[-1ex]{0pt}{3.5ex} $N$128 & 17.0 & 16.9 & 18.3 & 14.7 & 80 \\
\rule[-1ex]{0pt}{3.5ex} $N$1875 & 18.0 & 15.6 & 16.9 & 13.5 & 100 \\
\rule[-1ex]{0pt}{3.5ex} $N$191 & 17.7 & 16.9 & 18.3 & 15.0 & 120 \\
\rule[-1ex]{0pt}{3.5ex} $N$207 & 27.7 & 17.6 & 18.9 & 15.5 & 100 \\
\hline
\multicolumn{3}{@{}l@{}}{\hbox to 0pt{\parbox{120mm}{\footnotesize
\rule[-1ex]{0pt}{3.5ex} \footnotemark[$*$] All magnitudes are described in the AB magnitude system.

\rule[-1ex]{0pt}{3.5ex} \footnotemark[$\dagger$] For a point source with S$/$N $=$ 5 and $\phi$\timeform{1''.5} aperture.

\rule[-1ex]{0pt}{3.5ex} \footnotemark[$\ddagger$] Approximate exposure time required for background-limited performance (BLIP). We define BLIP as when the background noise becomes twice as large as the readout noise.
}\hss}}
    \end{tabular}
  \end{center}
\end{table*}

Throughput is defined as the ratio of the number of photons coming from an object into the Earth atmosphere of an aperture of the telescope to that of electrons detected, where atmospheric extinction is included.
The former is calculated with the effective collecting area of the telescope ($\sim$ 0.78 m$^{2}$) and the bandwidth of a filter used.
The latter is simply obtained by multiplying a count rate of an object (ADU s$^{-1}$) by the detector conversion factor.

\begin{figure}
  \begin{center}
    \FigureFile(80mm,80mm){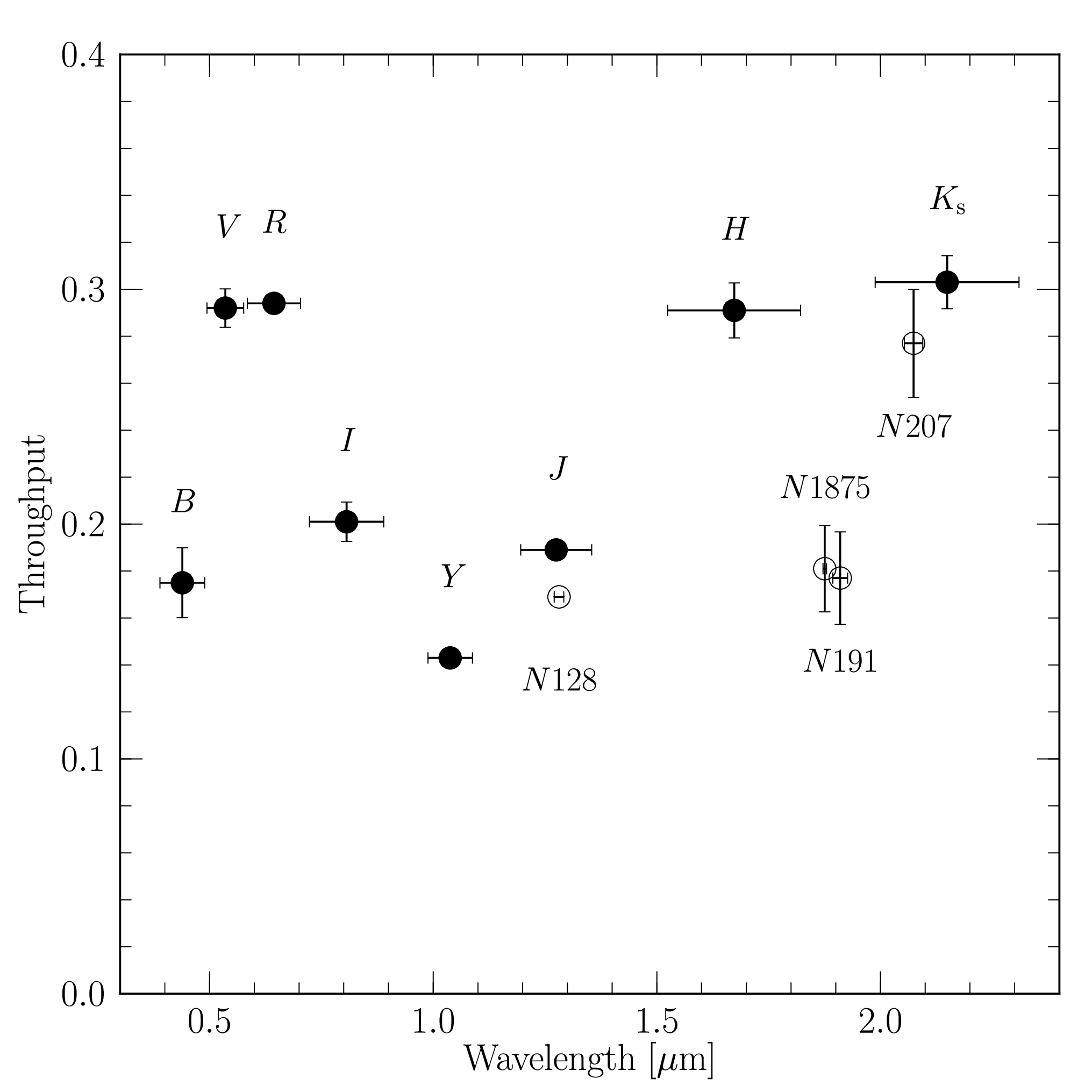}
  \end{center}
  \caption{Throughputs of the ANIR optical and NIR unit on the miniTAO telescope. The filled (open) circles show those of the broad-band (narrow-band) filters. The horizontal error bars represent the bandwidths, and the vertical error bars show 1$\sigma$ dispersions. Note that atmospheric extinction is not corrected.}
  \label{fig:efficiency}
\end{figure}

We use optical standard stars (\cite{Landolt92}) for the optical unit and 2MASS PSC stars within 11--17 mag and with a median photometric error of $\sim$ 0.1 mag for the NIR unit for the measurements.
The results are shown in Figure \ref{fig:efficiency} and summarized in Table \ref{tab:performance}.
Since the 2MASS catalog has only the $J$, $H$, and $K_{\rm{s}}$-band data, we calculate the magnitudes of the PSC stars for the $Y$-band and NB filters as follows.
For the $Y$-band, we linearly (in logarithmic scale) extrapolate the $J$-band (1.275 $\mu$m) magnitude to the $Y$-band (1.038 $\mu$m) with a slope between the $J$ and $H$-bands (1.673 $\mu$m) for each star.
For the $N$128 filter, we use the $J$-band magnitude.
The lower throughput of the $N$128 filter compared to the $J$-band is attributed to the combination use of the $J$-band filter as a thermal blocker for observations with the $N$128 filter.
We calculate magnitudes at the $N$1875, $N$191 and $N$207 filters by linearly (in logarithmic scale) interpolating their $H$ and $K_{\rm{s}}$-band (2.149 $\mu$m) magnitudes.
We examine the influence of stellar photospheric features within the wavelength ranges of interest on the interpolation. At first, we use $J-H$ and $H-K_{\rm{s}}$ color-color diagram to explore which kind of stars dominate the 2MASS PSC, and derive median colors of $J-H \sim 0.7$ and $H-K_{\rm{s}} \sim 0.2$, which correspond to intrinsic colors of K- to M-type stars (\cite{Bessell88}; \cite{Stead11}) or would be those of G- to K-type stars with dust extinctions of $A_{V} \sim$ 1--2 mag for data in low Galactic latitudes such as Galactic plane and Large/Small Magellanic Clouds (\cite{Imara07}; \cite{Dobashi09}).
Then, we evaluate the influence of spectral features of those medium-mass stars by using a stellar atmosphere model of \citet{Kurucz93} containing O- to M-type stars. We find no significant differences ($\lesssim$ 0.01 mag) between magnitudes derived by interpolation and those derived by direct integration of the model spectrum convolved with the transmission curve of the NB filter, although those stars have hydrogen absorption lines, of course including Pa$\alpha$. Therefore, the magnitudes at those NB filters we use hereafter are thought to be reliable enough for the performance verification of those NB filters and the site evaluation (Section \ref{sect:pwv}).
The resulting throughputs of the $N$1875 and $N$191 filters are significantly lower than those of the $H$ and $K_{\rm{s}}$-bands, and have larger dispersion by a factor of two, which are thought to be due to the lower atmospheric transmittance combined with the airmasses of the observations and temporal fluctuations in the atmospheric absorption mainly from water vapor (see Figure \ref{fig:atran_paa} for the transmittance with a PWV of 0.5 mm at an airmass of 1.0).
We evaluate the PWV content above the TAO site by using the throughputs of those filters in Section \ref{sect:pwv}.
The throughput of the $N$207 filter is consistent with that of the $K_{\rm{s}}$-band when taking into account the difference in their average filter transmittances: $\sim$ 78\% ($N$207) and  $\sim$ 92\% ($K_{\rm{s}}$).
It should be noted that, although there is several night-sky OH emission lines within the bandpasses of the NB filters (e.g., \cite{Rousselot00}), we consider their influence on our result to be small because we perform aperture photometry with \textit{local} sky subtraction for each source.

The limiting magnitudes for the optical and NIR units are estimated using the dark current, readout noise, throughputs obtained above, and sky brightness (\textit{photons} cm$^{-2}$ s$^{-1}$ $\mu$m$^{-1}$ arcsec$^{-2}$).
The sky brightness is measured on a ``sky'' frame made by combining unregistered subsequent frames with objects masked.
Table \ref{tab:performance} gives the estimated limiting magnitudes for a point source with S$/$N $=$ 5, $\phi$\timeform{1.''5} aperture, and two different exposure times (60 and 600 s).
Also listed is an approximate exposure time ($T_{\rm{BLIP}}$) required to achieve background-limited performance (BLIP) that means the S$/$N of data is dominated by the background noise.
We define BLIP as when the Poisson noise of the sky background becomes twice as large as the readout noise.

\subsection{Pa$\alpha$ Imaging Performance}

\begin{figure*}
  \begin{center}
    \FigureFile(170mm,170mm){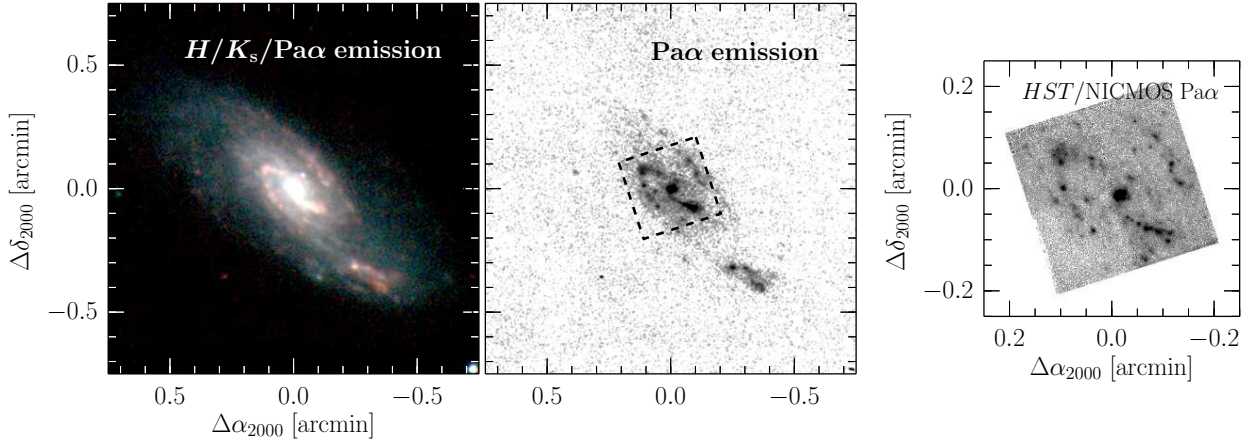}
  \end{center}
  \caption{Postage stamps of the nearby starburst galaxy, IC5179 ($\alpha$ = \timeform{22h16m09.10s}, $\delta$ = \timeform{-36d50m37.4s}, J2000.0), as an example demonstrating the ANIR Pa$\alpha$ imaging performance. The \textit{left} panel shows a \timeform{1.'5} $\times$ \timeform{1.'5} composite image (R: continuum-subtracted Pa$\alpha$, G: $K_{\rm{s}}$-band, B: $H$-band), oriented with north up and east to the left.  The \textit{middle} panel shows the continuum-subtracted Pa$\alpha$ image (on-source exposure time of 4200 s). For comparison, a Pa$\alpha$ emission line image taken by \textit{HST}/NICMOS is also shown in the \textit{right} panel (exposure time of 960 s) from which the F187N image is subtracted from the F190N image as a continuum emission. Note that the NICMOS image covers only approximately the central \timeform{0.'4} $\times$ \timeform{0.'4} region of the galaxy.}
  \label{fig:LIRG_paa}
\end{figure*}

Here we briefly demonstrate the performance of the miniTAO/ANIR \textit{ground-based} Pa$\alpha$ observations.
The reduced Pa$\alpha$ emission image of the nearby starburst galaxy, IC5179 taken with the $N$191 filter, is shown in Figure \ref{fig:LIRG_paa}, demonstrating that the TAO site indeed provides access to the Pa$\alpha$ emission line from the ground.
Star-forming regions associated with the spiral arm and the nucleus are clearly seen.
For a comparison of its sensitivity, the same galaxy taken by the \textit{HST}/NICMOS F190N filter is also shown in the right-hand panel of Figure \ref{fig:LIRG_paa}, from which the F187N image is subtracted as a continuum emission.
A quantitative comparison confirms the consistency of Pa$\alpha$ fluxes obtained with ANIR and NICMOS within an accuracy of 10\% (\cite{Tateuchi12a}).
We refer the reader to \citet{Tateuchi12a} and \citet{Tateuchi14} for the data reduction processes, verification, quantitative analyses, and comparison of the performance with NICMOS, of Pa$\alpha$ emission line data taken by the $N$1875 and $N$191 filters.

\section{Site Evaluation through Pa$\alpha$ Narrow-band Observations}
\label{sect:pwv}

Finally, we evaluate how the site is suitable for the infrared astronomy in terms of PWV by using the $N$1875 and $N$191 data taken between 2009 and 2011.
We note that our approach using the NB filters around 1.9 $\mu$m which are sensitive to PWV is independent of any previous studies carried out at (the vicinity of) the Chajnantor Plateau (e.g., \cite{Giovanelli01b}; \cite{Erasmus02}; \cite{Peterson03}; \cite{Tamura11}).

The throughputs derived in Section \ref{sect:throughput} (hereafter $\eta_{\rm{obs}}$) consist of the following factors: (i) the reflectivity of the telescope mirrors ($T_{\rm{Tel}}$), (ii) the system efficiency of ANIR excluding the filter transmittance ($T_{\rm{ANIR}}$), (iii) the filter transmittance ($T_{\rm{filter}}$), (iv) atmospheric transmittance ($T_{\rm{atm}}^{\rm{PWV}}$) at the zenith which is dependent on PWV.
Then, the throughput is described as:
\begin{eqnarray}
 \label{eq:eta_Paa}
  \eta_{\rm{obs}}^{\rm{band}}\ &=&\ T_{\rm{Tel}}^{\rm{band}} \times T_{\rm{ANIR}}^{\rm{band}} \times T_{\rm{filter}}^{\rm{band}} \times (T_{\rm{atm}}^{\rm{PWV,\ band}})^{X}
\end{eqnarray}
where the superscript ``band'' denotes that the factor depends on the wavelength measured and $X$ is the airmass which is necessary for considering the optical path length of the atmosphere during the observation.
For the sake of simplicity, we assume a homogeneous atmosphere in terms of airmass to obtain the atmospheric transmittance at a given airmass by scaling that at the airmass of 1.0.
In fact, such a simple scaling tends to overestimate the atmospheric extinction at $X$ $>$ 1 due to the non-linear dependence of the amount of extinction on airmass (\cite{Manduca79}; \cite{Tokunaga02}).
Thus, the resultant quantities, $T_{\rm{atm}}$ and PWV, are considered to be lower and upper limits, respectively.
$T_{\rm{filter}}$ is measured at 77 K in the laboratory (Table \ref{tab:spec_nir_NB}).
We calculate $T_{\rm{atm}}^{\rm{PWV}}$ using the ATRAN model.
As the atmospheric transmittances at the $H$ and $K_{\rm{s}}$-bands are largely independent regardless of PWV (see Figure \ref{fig:atran_nir}), the factor $T_{\rm{Tel}}^{\rm{band}} \times T_{\rm{ANIR}}^{\rm{band}}$ can be calculated, for example, for the $K_{\rm{s}}$-band as the following:
\begin{eqnarray}
 \label{eq:T_const}
  T_{\rm{Tel}}^{K_{\rm{s}}} \times T_{\rm{ANIR}}^{K_{\rm{s}}}  \ &=&\ \frac{\eta_{\rm{obs}}^{K_{\rm{s}}}}{T_{\rm{filter}}^{K_{\rm{s}}} \times (T_{\rm{atm}}^{K_{\rm{s}}})^{X}}
\end{eqnarray}
In the same way, we calculate the factor for the $H$-band ($T_{\rm{Tel}}^{H} \times T_{\rm{ANIR}}^{H}$).
Since the factor for the $H$-band is much the same as that for the $K_{\rm{s}}$-band with a dispersion of a few percent, indicating that the factor has small dependence on wavelength, we then interpolate $T_{\rm{Tel}}^{H} \times T_{\rm{ANIR}}^{H}$ and $T_{\rm{Tel}}^{K_{\rm{s}}} \times T_{\rm{ANIR}}^{K_{\rm{s}}}$ to estimate the same factor for a NB filter ($N$1875 or $N$191), $T_{\rm{Tel}}^{\rm{NB}} \times T_{\rm{ANIR}}^{\rm{NB}}$.
Finally, we obtain PWV by iteratively calculating the left-hand integral in Equation (\ref{eq:integral}) within the bandpass [$\lambda_{1}$, $\lambda_{2}$] of the NB filter with changing PWV to make it equal to the right-hand value.
\begin{eqnarray}
 \label{eq:integral}
  \frac{\int_{\lambda_{1}}^{\lambda_{2}} T_{\rm{filter}}^{\rm{NB}} \times (T_{\rm{atm}}^{\rm{PWV,\ NB}})^{X} d\lambda}{\int_{\lambda_{1}}^{\lambda_{2}} d\lambda}\ &=&\ \frac{\eta_{\rm{obs}}^{\rm{NB}}}{T_{\rm{Tel}}^{\rm{NB}} \times T_{\rm{ANIR}}^{\rm{NB}}}
\end{eqnarray}

\begin{figure*}
  \begin{center}
    \FigureFile(85mm,85mm){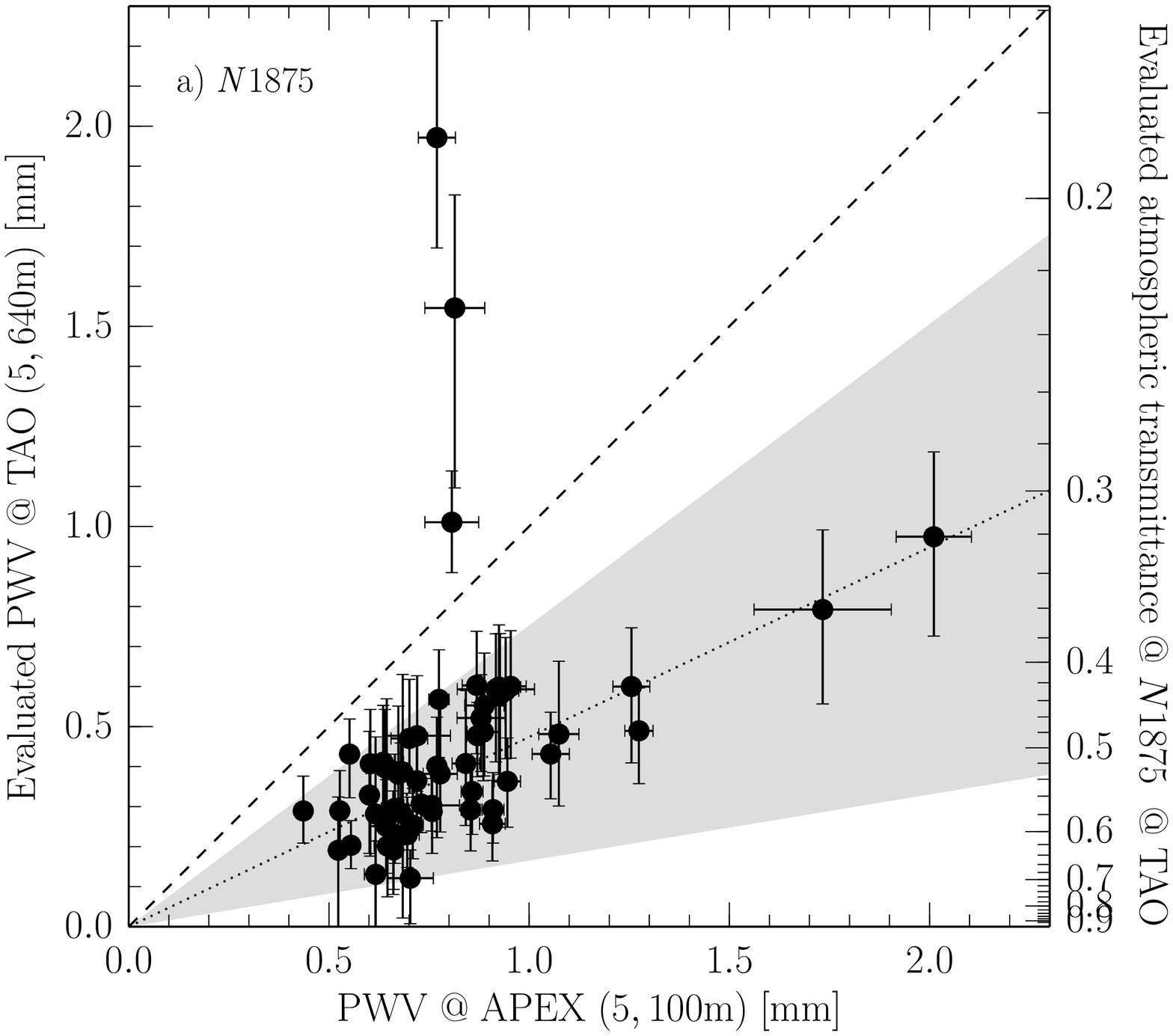}
    \FigureFile(85mm,85mm){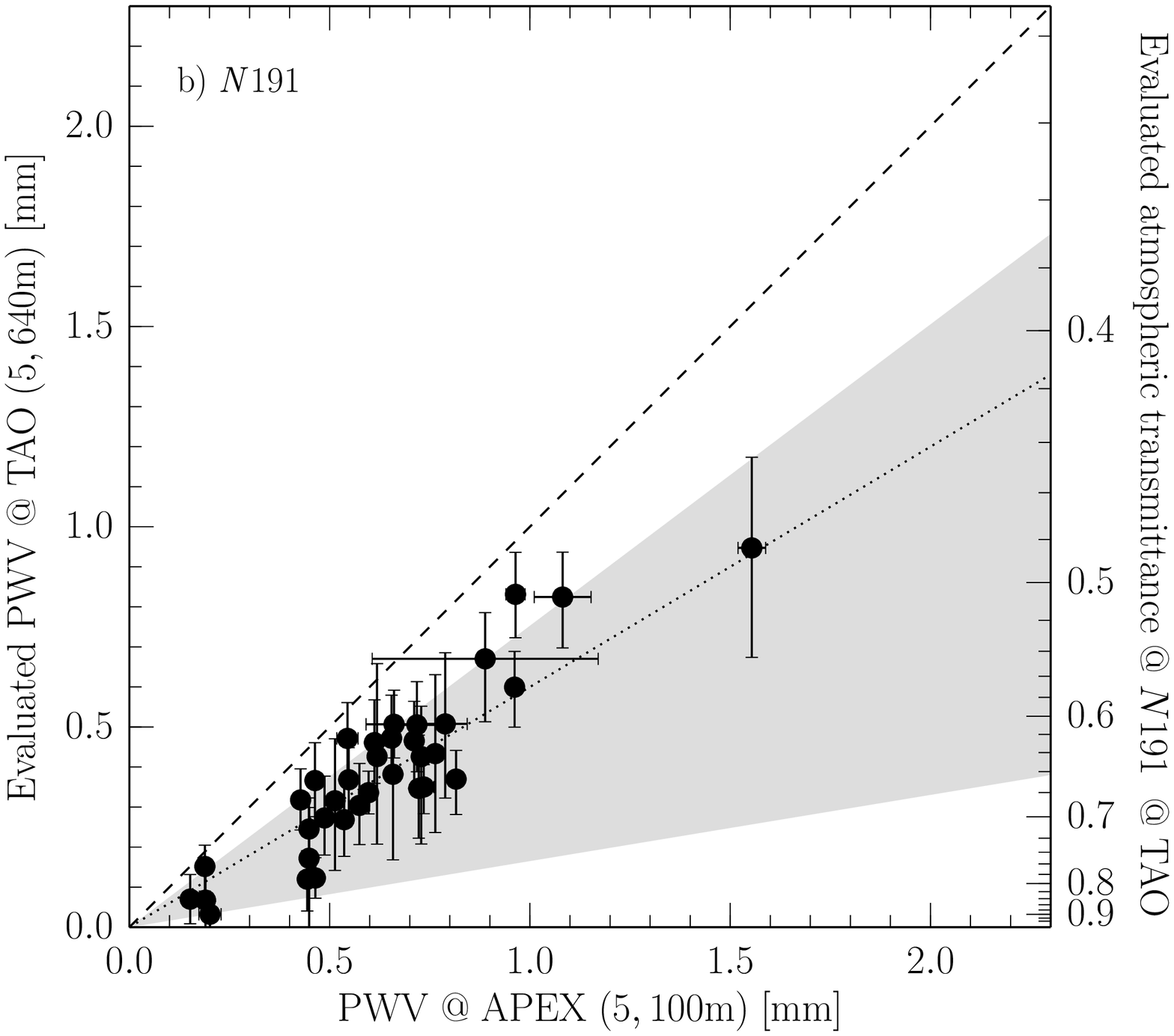}
  \end{center}
  \caption{\textit{Left}: Comparison of PWV values measured at the APEX site (5,100 m) with those at the TAO site (5,640 m) evaluated via the atmospheric transmittance at the $N$1875 filter. The vertical error bars account for only random errors and are the sum in quadrature of (i) the 1$\sigma$ uncertainties in the throughput of the $N$1875 filter which is used to estimate the atmospheric transmittance (the right-hand ordinate) and (ii) the frame-by-frame fluctuation of the throughput in the same dithering sequence (0.14 mm in PWV for all the data). The horizontal error bars indicate 1$\sigma$ dispersions of the APEX PWV values during $N$1875 observations for each object (typical integration time $\sim$ 1200 s including overheads). The dashed line indicates a one-to-one relation and the dotted line shows the best-fitting linear relation with a slope of 0.47. The \textit{shaded} region shows the possible region expected from the exponential distribution of the water vapor with a scale height of 0.3--1.9 km.
\textit{Right}: Same as the \textit{left-hand} panel, but for the $N$191 data. The best-fitting  (dotted) line has a slope of 0.60.}
  \label{fig:pwv_comp}
\end{figure*}

The evaluated PWV values and the corresponding atmospheric transmittances at the $N$1875 ($N$191) filter are shown in the left- and right-hand ordinates of Figure \ref{fig:pwv_comp}a (\ref{fig:pwv_comp}b), respectively.
Each data point is calculated by using a reduced and combined NB image of several dithering frames which has a total integration time of $\sim$ 1200 s.
The vertical error bars include in quadrature (i) the 1$\sigma$ uncertainty of the throughput of the NB data and (ii) frame-by-frame fluctuations of the throughput in the same dithering sequence.
We evaluate the latter factor by processing individual frames in the same manner as described above for two objects: one with a higher ($\sim$ 0.8 mm) and another with a lower ($\sim$ 0.2 mm) PWV evaluated from their stacked image. 
We find a frame-by-frame fluctuation of about 0.14 mm for both objects, and we thus introduce a constant factor of 0.14 mm as the frame-by-frame fluctuation for all the data in  Figure \ref{fig:pwv_comp}.
Our result tends to show lower PWV values (the median and its 1$\sigma$ dispersion are 0.40 $\pm$ 0.30 for $N$1875 and 0.37 $\pm$ 0.21 mm for $N$191) than those reported previously (i.e., \cite{Giovanelli01b}; \cite{Erasmus02}).
This may be partly because our observations have been carried out mostly in May and October, when the climate is expected to be drier than the yearly mean.
Note that the effect of the airmass has been corrected in Figure \ref{fig:pwv_comp}, so that the dispersion of each measurement of PWV is likely to be caused by the temporal fluctuation in the atmospheric absorption.

For a quantitative comparison of the result, we use archival PWV data measured by a radiometer of Atacama Pathfinder EXperiment (APEX, \cite{Gusten06}) located near the base of Cerro Chajnantor at an altitude of 5,100 m.
We extract the APEX PWV data during our $N$1875 ($N$191) observations, which are plotted along the abscissa in Figure \ref{fig:pwv_comp}a (\ref{fig:pwv_comp}b).
The error bars represent 1$\sigma$ dispersion of the extracted APEX PWV during our observations ($\sim$ 1200 s per object).
We clearly see that the PWV evaluated at the TAO site is remarkably lower (the median ratio and its 1$\sigma$ dispersion are 49\% $\pm$ 38\% for $N$1875 and 59\% $\pm$ 26\% for $N$191) than those measured at the APEX site.
A simple linear fit through the origin to the distribution yields a slope of 0.47 for $N$1875 and 0.60 for $N$191 (the \textit{dotted} line in Figure \ref{fig:pwv_comp}).

Let us now consider the vertical distribution of the water vapor to discuss possible causes of the difference in the PWV between the APEX (PWV$_{\rm{APEX}}$) and the TAO (PWV$_{\rm{TAO}}$) sites.
When the water vapor is assumed to be distributed exponentially, PWV is derived by integrating an exponential profile, $\rho\ =\ \rho_{\rm{0}}\ \rm{exp}[$$-(h-h_{0})/h_{e}$$]$, over altitude $h$, where $\rho$ is the water vapor density (in kg m$^{-3}$) at a given altitude $h$ above sea level (in km), $\rho_{0}$ the density at a reference altitude $h_{0}$, $h_{e}$ the scale height at which the water vapor density decreases by a factor of $e$.
\citet{Giovanelli01b} have measured the vertical distribution of the water vapor above the Chajnantor plateau by a combination of radiometric (at both 183 and 225 GHz) and radiosonde measurements, and derived the median scale height $h_{e}$ $\sim$ 1.13 km by fitting the individual distributions to the exponential profile with $h_{0}$ of 5.0 km (i.e., Chajnantor Plateau).
By using a scale height of 1.13 km, the PWV at the TAO site is calculated for a given PWV at the APEX site, as follows (see also \authorcite{Otarola10} \yearcite{Otarola10}, \yearcite{Otarola11}).
The density $\rho_{0}$ is found by integrating the above equation above 5,100 m since the integral equals to the PWV at the APEX site.
For example, the PWV at the APEX site of 1.0 mm leads to $\rho_{0} \sim$ 1.44 g cm$^{-3}$.
Then, the PWV at the TAO site is derived, to be $\sim$ 0.62 mm in the case of the example, by integrating the equation above 5,640 m with $\rho_{0}$ substituted.
Using the measurements of the exponential scale height at night time (the peak-to-peak values in $h_{e}$ of $\sim$ 0.3--1.9 km) by \citet{Giovanelli01b}, we derive the corresponding PWV values at the TAO site as a function of the PWV at the APEX site, which are shown in Figure \ref{fig:pwv_comp} with a \textit{shaded} region.
We find that almost all of the data points are in excellent agreement, within the uncertainties, with the expectation of such exponential distributions of the water vapor having the measured scale heights.
Note that there are a few outliers above the region, which might be caused by liquid phase of water (fog or clouds) in the atmosphere which is not detected by the APEX radiometer operating at 183 GHz, as suggested by previous studies (\cite{Matsushita03}; \cite{Tamura11}).
We should also note the possible influence of the presence of temperature inversion layers on the water vapor distribution.
\citet{Giovanelli01b} find that their radiosonde data often show temperature inversions which make the distribution far from the exponential shape, so that much of the water vapor would be trapped below the inversion layers.
To explore the influence requires a large sample of radiosonde data (or any equivalent data on the vertical distributions of the water vapor) and it is far beyond the scope of this paper, but we consider that our conclusion of the advantage of the site remains unrevised, because temperature inversions often take place below the altitude of 5,500 m at night time (\cite{Giovanelli01b}), indicating systematically lower PWV content at the TAO site than at the APEX site, which is qualitatively the same trend as derived from Figure \ref{fig:pwv_comp}.

While there is relatively large uncertainties in our analysis, it is encouraging to see that the low PWV content at the TAO site is confirmed by independent methods (i.e., NB imaging and radiosonde), which suggests that the TAO site at the summit of Cerro Chajnantor is suitable for infrared astronomy, and in particular that miniTAO/ANIR has an excellent capability for Pa$\alpha$ observations at around $\lambda=$ 1.9 $\mu$m.

\section{Summary}
We have developed a near-infrared camera, ANIR (Atacama NIR camera), for the 1.0-m miniTAO telescope installed at the summit of Cerro Chajnantor (an altitude of 5,640 m) in northern Chile.
The camera covers a field of view of \timeform{5'.1} $\times$ \timeform{5'.1} with a spatial resolution of \timeform{0''.298} pixel$^{-1}$ in the wavelength range of 0.95 to 2.4 $\mu$m.

The unique feature of the camera coupled with advantages of the site is a capability of narrow-band imaging observations of a strong hydrogen emission line, Paschen-$\alpha$ (Pa$\alpha$), at $\lambda=$ 1.8751 $\mu$m from the ground, at which wavelength it has been quite difficult to conduct ground-based observations so far due to deep atmospheric absorption mostly from water vapor.
We have been successfully obtaining Pa$\alpha$ images of Galactic objects and nearby galaxies since the first light observation with ANIR in 2009.

The throughputs at the narrow-band filters ($N$1875 and $N$191) show larger dispersion ($\sim$ 10\%) than those of the broad-band filters (a few percent), indicating that they are affected by temporal fluctuations in Precipitable Water Vapor (PWV) above the site.
Combining the atmospheric transmission model with the throughputs, we evaluated the atmospheric transmittance at the narrow-band filters and the PWV content at the site.
We find that the median and the dispersion of the PWV are 0.40 $\pm$ 0.30 for $N$1875 and 0.37 $\pm$ 0.21 mm for $N$191 data.
Comparing those data with the radiometer data taken by APEX, we find that the site has remarkably lower (49\% $\pm$ 38\% for $N$1875 and 59\% $\pm$ 26\% for $N$191) PWV values than the APEX site (5,100 m).
The differences in PWV between those sites are found to be in excellent agreement with those expected from the exponential distribution of the water vapor with scale height within 0.3--1.9 km which has been measured using the radiosonde at night time (\cite{Giovanelli01b}).
Although temperature inversions (which mostly take place below the summit at night-time) make the water vapor distribution far from exponential, it leads to lower PWV content above the inversion layers, which is qualitatively consistent with our findings described above.
Taken all together, we conclude that miniTAO/ANIR and the site, the summit of Cerro Chajnantor, provides us with an excellent capability for \textit{ground-based} Pa$\alpha$ observations.

\bigskip
\section*{Acknowledgments}
\noindent
We are extremely grateful to the anonymous referee for useful comments and suggestions that helped improve the quality of the paper.
%
We would like to acknowledge Mar\'{\i}a Teresa Ruiz Gonz\'{a}lez, Leonardo Bronfman, Mario Hamuy, and Ren\'{e} Alejandro M\'{e}ndez Bussard at the University of Chile for their support for the TAO project.
Operation of ANIR on the miniTAO telescope is supported by Ministry of Education, Culture, Sports, Science and Technology of Japan, Grant-in-Aid for Scientific Research (17104002, 20040003, 20041003, 21018003, 21018005, 21684006, 22253002, 22540258, and 23540261) from Japan Society for the Promotion of Science (JSPS).
Part of this work has been supported by the Institutional Program for Young Researcher Overseas Visits operated by JSPS.
Part of this work has been supported by the Advanced Technology Center, National astronomical observatory of Japan (NAOJ).
Part of this work has been supported by NAOJ Research Grant for Universities, and by Optical \& Near-Infrared Astronomy Inter-University Cooperation Program, supported by the MEXT of Japan.
Part of the ANIR development was supported by the Advanced Technology Center, NAOJ.
The PACE HAWAII-2 FPA array detector has been generously leased by Subaru Telescope, NAOJ.
The Image Reduction and Analysis Facility (IRAF) used in this paper is distributed by the National Optical Astronomy Observatories, which are operated by the Association of Universities for Research in Astronomy, Inc., under cooperative agreement with the National Science Foundation.
This publication makes use of data products from the Two Micron All Sky Survey, which is a joint project of the University of Massachusetts and the Infrared Processing and Analysis Center/California Institute of Technology, funded by the National Aeronautics and Space Administration and the National Science Foundation.
Some/all of the data presented in this paper were obtained from the Mikulski Archive for Space Telescopes (MAST). STScI is operated by the Association of Universities for Research in Astronomy, Inc., under NASA contract NAS5-26555. Support for MAST for non-\textit{HST} data is provided by the NASA Office of Space Science via grant NNX09AF08G and by other grants and contracts.
We would like to thank APEX for providing their radiometer data on the web.
APEX is a collaboration between the Max Planck Institute for Radio Astronomy, the European Southern Observatory, and the Onsala Space Observatory.



\begin{thebibliography}{}
\bibitem[Alonso-Herrero {\etal} (2006b)]{Alonso-Herrero06b} Alonso-Herrero,~A., Rieke,~G.~H., Rieke,~M.~J., Colina,~L., P{\'e}rez-Gonz{\'a}lez,~P.~G., \& Ryder,~S.~D. 2006b\ {\apj}, 650, 835
\bibitem[Bessell \& Brett (1988)]{Bessell88} Bessell,~M.~S. \& Brett,~J.~M., 1988, {\pasp}, 100, 1134
\bibitem[Bothwell {\etal} (2011)]{Bothwell11} Bothwell,~M.~S., {\etal}\ 2011, {\mnras}, 415, 1815
\bibitem[Dobashi {\etal} (2009)]{Dobashi09} Dobashi,~K., Bernard,~J.-P., Kawamura,~A., Egusa,~F., Hughes,~A., Paradis,~D., Bot,~C., \& Reach,~W.~T. 2009, {\aj}, 137, 5099
\bibitem[Dong {\etal} (2011)]{Dong11} Dong,~H., {\etal} 2011, {\mnras}, 417, 114
\bibitem[Erasmus \& Sarazin (2002)]{Erasmus02} Erasmus,~D., \& Sarazin,~M.\ 2002, ASPC, 266, 310E
\bibitem[Finger {\etal} (1998)]{Finger98} Finger,~G., Biereichel,~P., Mehrgan,~H., Meyer,~M., Moorwood,~A.~F., Nicolini,~G., \& Stegmeier,~J.\ 1998, Proc. SPIE, 3354, 87
\bibitem[Garc{\'{\i}}a-Mar{\'{\i}}n {\etal} (2006)]{Garcia06} Garc{\'{\i}}a-Mar{\'{\i}}n,~M., Colina,~L., Arribas,~S., Alonso-Herrero,~A., \& Mediavilla,~E. 2006, {\apj}, 650, 850
\bibitem[Giovanelli {\etal} (2001a)]{Giovanelli01a} Giovanelli,~R., {\etal}\ 2001a, {\pasp}, 113, 789
\bibitem[Giovanelli {\etal} (2001b)]{Giovanelli01b} Giovanelli,~R., {\etal}\ 2001b, {\pasp}, 113, 803
\bibitem[G{\"u}sten {\etal} (2006)]{Gusten06} G{\"u}sten,~R., Nyman,~L.~{\AA}., Schilke,~P., Menten,~K., Cesarsky,~C., \& Booth,~R.\ 2006, {\aap}, 454, L13
\bibitem[Hodapp {\etal} (1996)]{Hodapp96} Hodapp,~K.-W., {\etal}\ 1996, New Astron., 1, 177
\bibitem[Imara \& Blitz (2007)]{Imara07} Imara,~N. \& Blitz,~L. 2007, {\apj}, 662, 969 
\bibitem[Kerber {\etal} (2010)]{Kerber10} Kerber,~F., {\etal}\ 2010, Proc. SPIE, 7733, 77331M
\bibitem[Kennicutt (1998)]{Kennicutt98} Kennicutt,~Jr.,~R.~C.\ 1998, {\araa}, 36, 189
\bibitem[Komugi {\etal} (2012)]{Komugi12} Komugi,~S., {\etal}\ 2012, {\apj}, 757, 138
\bibitem[Kurucz (1993)]{Kurucz93} Kurucz,~R., 1993a, ATLAS9 Stellar Atmosphere Programs and 2 km/s grid. Kurucz CD-ROM No. 13. Cambridge, Mass.: Smithsonian Astrophysical Observatory, 1993., 13
\bibitem[Landolt (1992)]{Landolt92} Landolt,~A.~U.\ 1992, {\aj}, 104, 340
\bibitem[Lasker {\etal} (2008)]{Lasker08} Lasker,~B.~M., {\etal}\ 2008, {\aj}, 136, 735
\bibitem[Lord (1992)]{Lord92} Lord,~S.~D.\ 1992, NASA Technical Memorandum 103957
\bibitem[Manduca \& Bell (1979)]{Manduca79} {Manduca},~A. \& {Bell},~R.~A. 1979, {\pasp}, 91, 848
\bibitem[Matsushita \& Matsuo (2003)]{Matsushita03} Matsushita,~S. \& Matsuo,~H. 2003, {\pasj}, 55, 325
\bibitem[Mclean {\etal} (2012)]{Mclean12} McLean,~I.~S., Smith,~E.~C., Becklin,~E.~E., Dunham,~E.~W., Milburn,~J.~W., \& Savage,~M.~L.\ 2012, Proc. SPIE, 8446, 844619
\bibitem[Minezaki {\etal} (2010)]{Minezaki10} Minezaki,~T., {\etal}\ 2010, Proc. SPIE, 7733, 773356
\bibitem[Motohara {\etal} (2002)]{Motohara02} Motohara,~K., {\etal}\ 2002, PASJ, 54, 315
\bibitem[Motohara {\etal} (2008)]{Motohara08} Motohara,~K., {\etal}\ 2008, Proc. SPIE, 7012, 701244
\bibitem[Osterbrock (1989)]{Osterbrock89} Osterbrock,~D.~E. 1989, Astrophysics of Gaseous Nebulae and Active Galactic Nuclei (Mill Valley, CA: Univ. Science Books)
\bibitem[Ot{\'a}rola {\etal} (2010)]{Otarola10} {Ot{\'a}rola},~A., {\etal}\ 2010, {\pasp}, 122, 470
\bibitem[Ot{\'a}rola {\etal} (2011)]{Otarola11} Ot{\'a}rola,~A., Querel,~R., \& Kerber,~F. 2011, arXiv:1103.3025
\bibitem[Peterson {\etal} (2003)]{Peterson03} {Peterson},~J.~B., {Radford},~S.~J.~E., {Ade},~P.~A.~R., {Chamberlin},~R.~A., {O'Kelly},~M.~J., {Peterson},~K.~M. \& {Schartman},~E.\ 2003, {\pasp}, 115, 383
\bibitem[Piqueras L{\'o}pez {\etal} (2013)]{Piqueras13} Piqueras L{\'o}pez,~J., Colina,~L., Arribas,~S., \& Alonso-Herrero,~A.\ 2013, {\aap}, 553, A85
\bibitem[Pei (1992)]{Pei92} Pei,~Y.~C.\ 1992, {\apj}, 395, 130
\bibitem[Rousselot {\etal} (2000)]{Rousselot00} Rousselot,~P., Lidman,~C., Cuby,~J.-G., Moreels,~G. \& Monnet,~G. 2000, {\aap} 354, 1134
\bibitem[Sako {\etal} (2008a)]{Sako08a} Sako,~S., {\etal}\ 2008a, Proc. SPIE, 7012, 70122T
\bibitem[Sako {\etal} (2008b)]{Sako08b} Sako,~S., {\etal}\ 2008b, Proc. SPIE, 7021, 702128
\bibitem[Simons \& Tokunaga (2002)]{Simons02} Simons,~D.~A. \& Tokunaga,~A.\ 2002, {\pasp}, 114, 169
\bibitem[Skrutskie {\etal} (2006)]{Skrutskie06} Skrutskie,~M.~F., {\etal}\ 2006, {\aj}, 131, 1163
\bibitem[Stead \& Hoare (2011)]{Stead11} Stead,~J.~J. \& Hoare,~M.~G. 2011, {\mnras}, 418, 2219
\bibitem[Takeuchi {\etal} (2010)]{Takeuchi10} Takeuchi,~T.~T., Buat,~V., Heinis,~S., Giovannoli,~E., Yuan,~F.-T., Iglesias-P\'{a}ramo,~J., Murata,~K.~L., \& Burgarella,~D. 2010, {\aap}, 514, A4
\bibitem[Tamura {\etal} (2011)]{Tamura11} Tamura,~Y., {\etal}\ 2011, {\pasj}, 63, 347
\bibitem[Tateuchi {\etal} (2012a)]{Tateuchi12a} Tateuchi,~K., {\etal}\ 2012a, Proc. SPIE, 8446, 84467D
\bibitem[Tateuchi {\etal} (2012b)]{Tateuchi12b} Tateuchi,~K., {\etal}\ 2012b, Publ. Korean Astron. Soc., 27, 297
\bibitem[Tateuchi {\etal} (2014)]{Tateuchi14} Tateuchi,~K., {\etal}\ 2014, {\apjs}, accepted (arXiv:1412.3899)
\bibitem[Thompson {\etal} (1998)]{Thompson98} Thompson,~R.~I., Rieke,~M., Schneider,~G., Hines,~D.~C., \& Corbin,~M.~R. 1998, ApJL, 492, L95
\bibitem[Tokunaga {\etal} (2002)]{Tokunaga02} Tokunaga,~A.~T., Simons,~D.~A., \& Vacca, W. D.\ 2002, {\pasp}, 114, 180
\bibitem[Tokunaga \& Vacca (2005)]{Tokunaga05} Tokunaga,~A.~T \& Vacca,~W.~D. 2005, {\pasp}, 117, 421
\bibitem[Wang {\etal} (2010)]{Wang10} Wang,~Q.~D., {\etal}\ 2010, {\mnras}, 402, 895
\bibitem[Yoshii {\etal} (2010)]{Yoshii10} Yoshii,~Y., {\etal}\ 2010, Proc. SPIE, 7733, 773308
\bibitem[Yoshikawa {\etal} (2006)]{Yoshikawa06} Yoshikawa,~T., Omata,~K., Konishi,~M., Ichikawa,~T., Suzuki,~R., Tokoku,~C., Uchimoto,~Y.~K., \& Nishimura,~T. 2006, Proc. SPIE, 6274, 62740Y
\end{thebibliography}
\end{document}